\documentclass[11pt]{article}
\usepackage[T1]{fontenc}
\usepackage{lmodern}
\usepackage[a4paper,margin=1.85cm]{geometry}
\usepackage{amsmath,amssymb,bm,graphicx,booktabs,array,longtable}
\usepackage[colorlinks=true,linkcolor=blue,citecolor=blue,urlcolor=blue]{hyperref}
\hypersetup{pdftitle={Spectroscopy, Regge trajectories and decay properties of the bcqq and bqcq tetraquark systems},pdfauthor={Chetan Lodha and Ajay Kumar Rai}}
\newcommand{\Tbc}{bc\bar q\bar q}
\newcommand{\Tcomp}{bq\bar c\bar q}
\newcommand{\as}{\alpha_s}
\newcommand{\Sd}{S_d}
\newcommand{\Sa}{S_{\bar d}}
\begin{document}
\title{Spectroscopy, Regge trajectories and decay properties of the $bc\bar q\bar q$ and $bq\bar c\bar q$ tetraquark systems}
\author{Chetan Lodha \quad Ajay Kumar Rai\\[0.5em]
\small Department of Physics, Sardar Vallabhbhai National Institute of Technology, Surat, Gujarat 395007, India}
\date{}
\maketitle
\begin{abstract}
Bottom--charm tetraquarks probe the interplay of heavy-quark dynamics, light-quark correlations, color structure and nearby hadronic thresholds. We investigate the spectroscopy and strong-rearrangement properties of the $bc\bar q\bar q$ and $bq\bar c\bar q$ ($q=u,d$) systems in a Cornell-type diquark--antidiquark framework with nonrelativistic (NR) and semirelativistic (SR) kinetics. Scalar and axial-vector cluster configurations are treated explicitly, and model-parameter uncertainties are propagated into state-dependent mass errors. For the $\bar{\mathbf 3}_c\otimes\mathbf 3_c$ axial--axial ground states, the $0^+$ masses are $7530.1\pm75.4$ MeV (NR) and $7720.7\pm78.8$ MeV (SR) for $bc\bar q\bar q$, and $7307.6\pm90.5$ MeV (NR) and $7523.5\pm93.6$ MeV (SR) for $bq\bar c\bar q$. The multiplets follow $0^+<1^+<2^+$, and radial spacings decrease with excitation. In the $bq\bar c\bar q$ sector several low-lying configurations approach the $B^*D^*$ threshold: the axial--axial $0^+$ state lies $25.7$ MeV below it in NR but $190.2$ MeV above it in SR, so its threshold assignment is model sensitive and not evidence of robust strong-interaction stability. Spin and color wavefunctions are recoupled independently, giving configuration-dependent $PP$, $PV$, $VP$ and $VV$ rearrangement strengths and distinct color-singlet projections for the $\bar{\mathbf 3}_c\otimes\mathbf 3_c$ and $\mathbf 6_c\otimes\bar{\mathbf 6}_c$ sectors. The strong short-distance attraction of the sextet coefficient $k_s=-10/3$ makes the finite SR expansion unreliable in that branch, so quantitative masses are restricted to the antitriplet sector. The results provide experimentally useful signatures and a baseline for coupled-channel and color-mixing studies.
\end{abstract}

\noindent\textbf{Keywords:} Tetraquarks; Bottom--charm; Diquark--antidiquark model; Regge trajectories; Rearrangement decays.

\raggedbottom

\section{Introduction}
The spectroscopy of multiquark hadrons provides a means of studying how
confinement and spin-dependent interactions organize color-singlet systems
beyond conventional mesons and baryons. The quark-model classification and
the early treatment of multiquark configurations established the basic
setting for such investigations~\cite{GellMann,DeRujula,Jaffe}. The study
of heavy exotic hadrons has since brought compact configurations, hadronic
molecules and threshold effects into a common phenomenological
discussion~\cite{BrambillaReview,GuoMolecules,OlsenReview}. In this setting,
the mass spectrum is most informative when considered together with the
state's quantum numbers and accessible decay channels.

Tetraquarks containing bottom and charm constituents are particularly
interesting because the heavy masses reduce part of the kinetic energy,
while the light-quark degrees of freedom retain a substantial role in
the binding and spin structure. The two flavor sectors considered in this
work, $\Tbc$ and $\Tcomp$, illustrate different consequences of this
combination. The first contains a $bc$ diquark and can rearrange into
bottom and charmed mesons. The second contains a $bq$ diquark and a
$\bar c\bar q$ antidiquark; in addition to open-heavy final states,
it admits a $b\bar c$ meson accompanied by a light meson. Their conserved
charm quantum numbers are different, so the two systems represent distinct
flavor sectors rather than alternative clusterings of one state.

Several theoretical approaches have addressed mixed-heavy tetraquarks.
QCD sum rules employ suitable interpolating currents to examine masses
and decay couplings~\cite{ChenQSR,AgaevTbc,TangQSR}. Relativistic diquark
models and constituent-quark calculations investigate compact structures
with different treatments of confinement and short-range
interactions~\cite{EbertRel,ParkCQM,YangDH,DengDH,LuRel}.
Heavy-diquark symmetry, extended four-body bases and resonance methods
provide complementary descriptions~\cite{ChengHADS,ZhangHADS2026,NohComplete,MengRes}.
Chromomagnetic and mass-splitting models emphasize the dependence on
spin--color couplings~\cite{GuoCMI,LiSplit}, while bound-state, resonance
and diffusion Monte Carlo calculations examine the role of spatial
correlations~\cite{WuBound,Mutuk}. Differences among these predictions
reflect both the assumed interactions and the dynamical spaces in which
they are solved.

Lattice QCD gives an especially useful comparison near open-heavy
thresholds, and recent reviews summarize the increasing emphasis on scattering
information in addition to finite-volume energy levels~\cite{FrancisLattice}.
Isoscalar axial-vector calculations with two-meson and
diquark--antidiquark operators~\cite{Padmanath}, finite-volume scalar and
axial scattering analyses~\cite{Alexandrou}, and an isoscalar scalar
study~\cite{RadhakrishnanScalar} explore states that cannot be characterized
by a compact mass alone. Effective-field-theory analyses of the lattice
spectra further relate shallow states to their heavy-quark-spin
partners~\cite{Abolnikov}. Chiral effective-field-theory calculations of
$D^{(*)}\bar B^{(*)}$ interactions further find repulsion in the $I=1$
channels and attraction in the $I=0$ channels~\cite{LiuChEFT}. Comparisons
with these results must therefore match flavor and isospin as well as $J^P$,
and distinguish an isolated cluster energy from a pole of the scattering
amplitude.

The Cornell potential offers a convenient framework for organizing the
compact spectrum through its short-range Coulomb interaction and
long-range linear confinement~\cite{EichtenModel,EichtenComparison}.
Relativized potential models and systematic accounts of quark-bound-state
dynamics provide the corresponding context for kinetic and spin-dependent
terms~\cite{GodfreyIsgur,Lucha}. A two-stage diquark construction makes
the effects of heavy--heavy, heavy--light and light--light clustering
explicit and permits a common comparison of radial, orbital and spin
excitations. Related studies of charm mesons and singly charmed
tetraquarks illustrate this connection between conventional calibration
and multiquark spectroscopy~\cite{LodhaCharm}.

In the present work, we examine the NR and SR spectra of both bottom--charm
systems in the antitriplet and sextet color configurations. Scalar and
axial cluster assignments are retained explicitly. The main discussion
follows the first three $S$-wave, first two $P$-wave and first $D$-wave
multiplets; the complete three-radial-level spectra for all three orbital
sectors are given in the appendix. The purpose is to connect the resulting
mass hierarchy with threshold placement, Regge organization and the spin
structure of rearrangement decays. Particular attention is paid to the
antitriplet $\Tcomp$ scalar, whose relation to $B^*D^*$ changes between
the two prescriptions while lower strong channels remain accessible.

The distinctive scope of this study is the simultaneous treatment of two
conserved-flavor bottom--charm sectors within one calibrated framework, with
scalar and axial cluster spins kept explicit, six-parameter sensitivity
propagated through the spin-dependent spectrum, and spin and color recoupling
separated for the $\bar3_c\otimes3_c$ and $6_c\otimes\bar6_c$ bases. This
allows threshold placement, light-isospin selection and internal spin
structure to be compared on the same footing while keeping exact recoupling
statements separate from the sextet spectroscopy that lies outside the
controlled SR domain.

Section~\ref{sec:model} presents the theoretical framework, calibration
and spin basis. Section~\ref{sec:results} discusses spectroscopy,
threshold sensitivity and Regge behavior. Rearrangement channels and
experimental signatures are considered in Sec.~\ref{sec:threshold}.
Section~\ref{sec:discussion} discusses the two flavor systems separately
and compares their physical implications with other approaches.
The conclusions are given in Sec.~\ref{sec:conclusion}.

\section{Theoretical framework, calibration and spin basis}\label{sec:model}
\subsection{Color structure and Cornell dynamics}
The tetraquark is described as a diquark and an antidiquark in their
internal $S$-wave states, with relative orbital angular momentum $L$.
The configuration basis is written as
\begin{equation}
 \left|[(q_1q_2)_{R_d,\Sd}(\bar q_3\bar q_4)_{\bar R_d,\Sa}]_S,L;JM\right\rangle.
 \label{eq:basis}
\end{equation}
The cluster spins obey $\Sd,\Sa=0,1$, and angular-momentum addition gives
\begin{equation}
 |\Sd-\Sa|\leq S\leq\Sd+\Sa,\qquad |L-S|\leq J\leq L+S.
 \label{eq:angular}
\end{equation}
The parity is $P=(-1)^L$. Thus $S$, $P$ and $D$ intercluster waves have
positive, negative and positive parity, respectively. We use the notation
$n^{2S+1}L_J$, with $n=1$ for the lowest radial level. A charge-conjugation
quantum number is not assigned to these open-flavor systems.

The central interaction between constituents or effective clusters is
\begin{equation}
 V_C(r)=V_V(r)+V_S(r)+V_0
       =\frac{k_s\as}{r}+br+V_0,
 \label{eq:cornell}
\end{equation}
where $b$ is the confinement strength and $V_0$ fixes the additive energy
convention. The color factor follows from the quadratic Casimirs,
\begin{equation}
 \langle\bm F_1\!\cdot\!\bm F_2\rangle
 =\frac12\left[C_2(R)-C_2(R_1)-C_2(R_2)\right].
\end{equation}
For a $qq$ pair, $k_s=-2/3$ in $\bar3_c$ and $k_s=+1/3$ in $6_c$.
For the corresponding color-singlet diquark--antidiquark combinations,
\begin{equation}
 k_s(\bar3_c\otimes3_c)=-\frac43,\qquad
 k_s(6_c\otimes\bar6_c)=-\frac{10}{3}.
 \label{eq:color}
\end{equation}
The stronger intercluster Coulomb factor in the sextet channel competes
with the different internal cluster masses and confinement parameters.
Its sign alone therefore does not determine which complete tetraquark
configuration has the lower mass.

The NR Hamiltonian is
\begin{equation}
 H_{\rm NR}=m_d+m_{\bar d}+\frac{\bm p^2}{2\mu}+V_C(r)+V_{\rm SD},
 \qquad \mu=\frac{m_dm_{\bar d}}{m_d+m_{\bar d}},
 \label{eq:H}
\end{equation}
where $V_{\rm SD}$ contains the spin-dependent interactions. The radial
equation for the spin-independent energy $E_{nL}$ is
\begin{equation}
 \left[-\frac{1}{2\mu}\frac{d^2}{dr^2}
 +\frac{L(L+1)}{2\mu r^2}+V_C(r)\right]u_{nL}(r)=E_{nL}u_{nL}(r).
 \label{eq:radial}
\end{equation}
Regularity at the origin and decay at large $r$ select the radial
solutions, normalized by $\int_0^\infty |u_{nL}(r)|^2dr=1$.
We expand $u_{nL}(r)$ in $r^{L+1}\exp(-\nu_i r^2)$ with 48
geometrically spaced ranges from $0.008$ to $40$ GeV$^{-1}$.
The generalized eigenvalue problem is solved after removing overlap
eigenvalues below $10^{-12}$. Increasing to 60 ranges over
$0.004$--$50$ GeV$^{-1}$ changes every reported mass by less
than $0.001$ MeV. Refined radial shooting provides an independent
check of representative antitriplet and sextet central energies.
The spin terms are evaluated on the normalized radial eigenfunctions.

The semirelativistic motivation is the expansion
\begin{align}
 T_{\sqrt{}}&=\sum_{i=d,\bar d}\left[\sqrt{\bm p^2+m_i^2}-m_i\right],\label{eq:sqrt}\\
 T_{\sqrt{}}&=\frac{p^2}{2\mu}-\frac18\sum_i\frac{p^4}{m_i^3}
             +\frac1{16}\sum_i\frac{p^6}{m_i^5}+\cdots .\label{eq:series}
\end{align}
In the effective kinetic prescription associated with the spectra, a
characteristic momentum $Q^2=\langle\bm p^2\rangle$ is evaluated from the spin-independent
radial solution and used in a finite series,
\begin{equation}
 \delta T_K=\sum_{k=2}^{K}\binom{1/2}{k}Q^{2k}
       \left(m_d^{1-2k}+m_{\bar d}^{1-2k}\right).
 \label{eq:effectivekinetic}
\end{equation}
We retain terms through $K=5$. The additive constant $V_0$ shifts the
energy but does not enter $Q^2$. This is an effective
momentum-scale approximation: $Q^{2k}$ is not the independently evaluated
moment $\langle p^{2k}\rangle$. Consequently, the SR masses are used as
an alternative model estimate, without identifying them with an exact
solution of Eq.~\eqref{eq:sqrt}. The NR and SR parameter sets and internal
cluster masses also differ, so their full mass displacement includes
calibration effects as well as the kinetic prescription.

Inverse-mass interactions are relevant in systematic heavy-quark
potential expansions~\cite{BrambillaPNRQCD,PinedaVairo}. The commonly
considered correction
\begin{equation}
 V^{(1)}(r)=-\left(\frac1{m_d}+\frac1{m_{\bar d}}\right)
             \frac{C_FC_A\as^2}{4r^2},
 \label{eq:inverse}
\end{equation}
with $C_F=4/3$ and $C_A=3$, is evaluated separately as a diagnostic in the associated numerical
formulation; it is not part of the central potential in
Eq.~\eqref{eq:radial} or the displayed mass sum. This distinction avoids
treating an unregularized attractive $1/r^2$ term as part of the solved
Cornell Hamiltonian.

\subsection{Spin interactions and light-flavor assignments}
The Gaussian-smeared contact term is
\begin{equation}
 V_{SS}=-\frac{8\pi k_s\as}{3m_dm_{\bar d}}
 \left(\frac{\sigma}{\sqrt\pi}\right)^3e^{-\sigma^2r^2}
 \bm S_d\cdot\bm S_{\bar d},
 \label{eq:ss}
\end{equation}
with
\begin{equation}
 \langle\bm S_d\cdot\bm S_{\bar d}\rangle
 =\frac{S(S+1)-\Sd(\Sd+1)-\Sa(\Sa+1)}2.
 \label{eq:dot}
\end{equation}
For two axial clusters, the values are $-2,-1,+1$ for $S=0,1,2$.
If either cluster is scalar, the intercluster contact term vanishes.
The diagonal spin--orbit prescription is
\begin{equation}
 V_{LS}=\left[-\frac{3k_s\as}{2m_dm_{\bar d}r^3}
              -\frac{b}{2m_dm_{\bar d}r}\right]\bm L\cdot\bm S,
 \label{eq:ls}
\end{equation}
where
$\langle\bm L\cdot\bm S\rangle=[J(J+1)-L(L+1)-S(S+1)]/2$.
In the equal-mass limit, Eq.~\eqref{eq:ls} has the standard
$(3V_V'-V_S')/(2m^2r)$ form for vector Coulomb and scalar confinement
interactions~\cite{Lucha}. We define the tensor operator as
$S_{12}=4[3(\bm S_d\!\cdot\!\hat r)(\bm S_{\bar d}\!\cdot\!\hat r)
-\bm S_d\!\cdot\!\bm S_{\bar d}]$, with
$V_T=-k_s\as S_{12}/(4m_dm_{\bar d}r^3)$.
For two axial clusters its diagonal matrix element is
\begin{equation}
 \langle S_{12}\rangle=-f_S
 \frac{3x^2+\tfrac32x-L(L+1)S(S+1)}{(2L-1)(2L+3)},
 \quad x=\langle\bm L\!\cdot\!\bm S\rangle,
 \label{eq:tensor}
\end{equation}
where $f_1=8$ and $f_2=8/3$. It vanishes for $S=0$, $L=0$
or a scalar cluster. Explicit spin coupling and angular integration
verify this normalization. All finite $P$- and $D$-wave radial
matrix elements are integrated to the origin.

The mass of a level is organized as
\begin{equation}
 M(n^{2S+1}L_J)=M^{(0)}_{nL}
             +\langle V_{SS}\rangle+\langle V_{LS}\rangle
             +\langle V_T\rangle,
 \label{eq:mass}
\end{equation}
where $M^{(0)}_{nL}$ includes the appropriate NR or SR spin-independent
estimate. The $P$- and $D$-wave labels describe diagonal configurations.
Off-diagonal tensor, antisymmetric spin--orbit and color mixing are
beyond this approximation. Their evaluation requires a common coupled
basis, for which few-body expansion methods offer a natural
extension~\cite{Hiyama}.

Fermi symmetry constrains the light antidiquark in $\Tbc$. For an
internally symmetric spatial wavefunction, the color, spin and flavor
product must be antisymmetric. A color-$3_c$ antidiquark permits
$(I,\Sa)=(1,1)$ or $(0,0)$, whereas a color-$\bar6_c$ antidiquark
permits $(I,\Sa)=(1,0)$ or $(0,1)$. We restrict the numerical
$\Tbc$ spectrum to $I=1$: its antitriplet configurations are
$(\Sd,\Sa)=(0,1),(1,1)$ and its sextet configurations are
$(0,0),(1,0)$. The isoscalar configurations require a separate
calculation. For $\Tcomp$, both clusters contain unlike flavors and
all four scalar--axial combinations are retained in either color.
Its compact masses are taken to be isospin independent at this level;
the $I=0$ and $I=1$ decay channels are treated separately.

\subsection{Calibration and parameter-sensitivity errors}\label{sec:data}
The calculation uses fixed central parameter sets calibrated in the
conventional $D$, $B$ and $B_c$ sectors. Table~\ref{tab:calibration}
collects these NR and SR inputs, while Table~\ref{tab:mesons} lists central
ground-state checks. No parameter refit is performed. External daughter
masses used for threshold comparisons are specified independently in
Sec.~\ref{sec:threshold}.
% BEGIN INLINE tables/calibration
\begin{table}[!tbp]
\caption{Conventional-sector parameter sets retained for the model calibration. Masses, $\sigma$ and $V_0$ are in GeV; $b$ is in GeV$^2$. These are the central calibration inputs; tetraquark mass errors are propagated separately.}\label{tab:calibration}
\centering\small\setlength{\tabcolsep}{3.3pt}
\begin{tabular}{llrrrrrr}\toprule
Scheme & Sector & $m_1$ & $m_2$ & $\alpha_s$ & $b$ & $\sigma$ & $V_0$\\\midrule
NR & $D(c\bar q)$ & 1.480 & 0.355 & 0.675 & 0.1070 & 0.875 & -0.15\\
   & $B(b\bar q)$ & 4.950 & 0.355 & 0.600 & 0.1080 & 0.875 & -0.28\\
   & $B_c(b\bar c)$ & 4.950 & 1.480 & 0.450 & 0.1425 & 1.200 & -0.17\\
SR & $D(c\bar q)$ & 1.475 & 0.375 & 0.650 & 0.1070 & 0.875 & -0.14\\
   & $B(b\bar q)$ & 4.980 & 0.375 & 0.600 & 0.1070 & 0.875 & -0.30\\
   & $B_c(b\bar c)$ & 4.980 & 1.475 & 0.450 & 0.1450 & 1.200 & -0.19\\
\bottomrule\end{tabular}\end{table}

% END INLINE tables/calibration

The two-stage construction first generates scalar and axial diquark masses,
which are then used as effective constituents of the tetraquark Hamiltonian.
Table~\ref{tab:diquarks} lists the central reference values. The antitriplet
diquark has an attractive internal Coulomb term, whereas the sextet internal
term is repulsive. The subsequent intercluster interaction is a separate
step and its color coefficient must therefore not be inferred from the
ordering of the isolated cluster masses. Conventional beauty spectroscopy
provides another application of this calibration strategy~\cite{LodhaBeauty}.
% BEGIN INLINE tables/mesons
\begin{table}[!tbp]
\caption{Calculated central ground-state meson masses in MeV. A dagger marks an effective SR kinetic series with $Q^2/m_{\min}^2\ge1$. These values are calibration checks; tetraquark parameter errors are propagated at the tetraquark stage.}\label{tab:mesons}
\centering\small\setlength{\tabcolsep}{5pt}
\begin{tabular}{lcrr}\toprule
Meson & $J^P$ & NR & SR\\\midrule
$D$ & $0^-$ & 1870.8 & $1879.5^{\dagger}$\\
$D^*$ & $1^-$ & 2006.6 & $2006.8^{\dagger}$\\
$B$ & $0^-$ & 5280.0 & $5295.1^{\dagger}$\\
$B^*$ & $1^-$ & 5319.7 & $5334.8^{\dagger}$\\
$B_c$ & $0^-$ & 6274.3 & 6257.2\\
$B_c^*$ & $1^-$ & 6313.2 & 6296.1\\
\bottomrule\end{tabular}\end{table}

% END INLINE tables/mesons

For $cq$, $bq$ and $bc$ diquarks, the quark masses, coupling and smearing
width are those of the corresponding meson input set, while the diquark
confinement strength is $b_{\rm meson}/4$ in NR and $b_{\rm meson}/2$ in SR.
The $qq$ inputs $(m_q,\alpha_s,b,\sigma)$ are
$(0.355,0.94,0.0325,0.95)$ in NR and
$(0.375,0.92,0.0625,0.9275)$ in SR, with masses and widths in GeV and
$b$ in GeV$^2$. At the tetraquark stage the intercluster central values are
$\alpha_s=0.45$, $\sigma=1.2$ GeV and $V_0=0$. The reference confinement
strengths are $b=0.1425$ GeV$^2$ (NR) and $0.1450$ GeV$^2$ (SR) in the
antitriplet branch.

The quoted mass errors are obtained from a simultaneous local sensitivity
analysis of the six model inputs $m_1$, $m_2$, $\alpha_s$, $\sigma$, $b$
and $V_0$ about their reference values. The symmetric variation ranges are
$\pm0.06$ GeV for $m_1$ and $m_2$, $\pm0.06$ for $\alpha_s$,
$\pm0.06$ GeV for $\sigma$, $\pm0.006$ GeV$^2$ for $b$, and
$\pm0.06$ GeV for $V_0$. These ranges define a uniform local sensitivity
test and are not interpreted as experimental uncertainties on the model
parameters. All six inputs are varied together over these domains. For each
state, the tabulated $M$ is the average mass over the resulting
multidimensional sensitivity set and $\Delta M$ is the corresponding
parameter-induced error.

For every parameter choice, the radial energy and wave function,
$\langle V_V\rangle$, $\langle V_S\rangle$, smeared spin--spin,
spin--orbit and tensor terms, the diagnostic $V^{(1)}$, relativistic kinetic
corrections, $Q^2$, wave-function quantities and normalization are evaluated
consistently before the final NR and SR masses are formed. Thus $\Delta M$
is state dependent and contains the response of the spin-dependent terms; it
is not attached afterward as a common error to a spin-averaged level. These
errors quantify sensitivity within the stated parameter domains and are not
statistical confidence intervals or a complete theory uncertainty.
Independent checks against the direct radial solver differ by less than
$0.04$ MeV for representative antitriplet points, well below the quoted mass
errors.

The variation ranges are absolute rather than relative, so they correspond to
different fractional changes of the individual inputs (about 1\% for
$m_1\sim6.5$ GeV in $bc\bar q\bar q$ but about 6\% for a light cluster mass
$m_2\sim1$ GeV), and correlations among the inputs induced by the meson
calibration are not included. The additive constant $V_0=0$ at the tetraquark
stage is not calibrated against any tetraquark datum, and its variation enters
$\Delta M$ directly. Errors are quoted to 0.1 MeV only so that the tables are
reproducible; given the nature of the sensitivity test, only the integer-MeV
values should be regarded as significant.

The color-singlet intercluster factors are fixed by Eq.~\eqref{eq:color}:
$k_s=-4/3$ for $\bar3_c\otimes3_c$ and $k_s=-10/3$ for
$6_c\otimes\bar6_c$. The strong sextet attraction drives representative
ground-state solutions to momenta for which the finite SR series violates
its convergence condition $Q^2/m_{\min}^2<1$. Accordingly, the quantitative
mass tables and phenomenological threshold analysis below are restricted to
the antitriplet branch. The sextet is retained as a color-structure and
recoupling diagnostic until a finite-size or fully relativistic treatment is
used.
% BEGIN INLINE tables/diquarks
\begin{table*}[!tbp]
\caption{Central scalar and axial diquark masses in MeV used as effective constituents of the tetraquark Hamiltonian. The light $ud$ entries require the flavor symmetry allowed by each color--spin combination; only the isovector light-antidiquark configurations enter the $bc\bar q\bar q$ spectrum. The resulting effective cluster masses $m_1$ and $m_2$ are included in the tetraquark error propagation.}\label{tab:diquarks}
\centering\small\setlength{\tabcolsep}{5pt}
\begin{tabular}{llrrrr}\toprule
Pair & Color & NR $S_d=0$ & NR $S_d=1$ & SR $S_d=0$ & SR $S_d=1$\\\midrule
$ud$ & $\bar3_c$ & 851.4 & 934.0 & 984.5 & 1101.0\\
$ud$ & $6_c$ & 1111.0 & 1099.4 & 1335.0 & 1312.9\\
$bc$ & $\bar3_c$ & 6486.3 & 6492.3 & 6592.3 & 6600.6\\
$bc$ & $6_c$ & 6672.7 & 6672.1 & 6827.6 & 6826.4\\
$bq$ & $\bar3_c$ & 5443.5 & 5449.0 & 5576.4 & 5584.9\\
$bq$ & $6_c$ & 5588.1 & 5587.3 & 5772.9 & 5771.2\\
$cq$ & $\bar3_c$ & 1967.4 & 1985.9 & 2077.4 & 2104.3\\
$cq$ & $6_c$ & 2132.9 & 2130.3 & 2293.8 & 2288.5\\
\bottomrule\end{tabular}\end{table*}

% END INLINE tables/diquarks

\section{Spectroscopy, threshold sensitivity and Regge organization}\label{sec:results}
\subsection{Antitriplet spectrum and cluster-spin dependence}
The antitriplet spectrum provides the quantitative
reference for comparing the two flavor sectors. Table~\ref{tab:ground}
collects the ground scalar and axial cluster configurations, while
Table~\ref{tab:spectrum} gives the axial--axial radial and orbital
multiplets. All masses in these tables are quoted as $M\pm\Delta M$, where $\Delta M$ is the parameter-sensitivity error defined in Sec.~\ref{sec:data}.
% BEGIN INLINE tables/ground
\begin{table*}[!tbp]
\caption{Ground-state antitriplet masses in MeV. Each entry is quoted as $M\pm\Delta M$, where $\Delta M$ is the propagated parameter-sensitivity error. The $bc\bar q\bar q$ spectrum has $I=1$; $bq\bar c\bar q$ compact masses are isospin independent at this level.}\label{tab:ground}
\centering\small\setlength{\tabcolsep}{4pt}
\begin{tabular}{llccrr}\toprule
System & Color & $(S_d,S_{\bar d})$ & $J^P$ & $M_{\rm NR}\pm\Delta M_{\rm NR}$ & $M_{\rm SR}\pm\Delta M_{\rm SR}$\\\midrule
$bc\bar q\bar q$ & $\bar3_c\otimes3_c$ & $(0,1)$ & $1^{+}$ & $7587.1\pm69.8$ & $7775.6\pm72.2$\\
$bc\bar q\bar q$ & $\bar3_c\otimes3_c$ & $(1,1)$ & $0^{+}$ & $7530.1\pm75.4$ & $7720.7\pm78.8$\\
$bc\bar q\bar q$ & $\bar3_c\otimes3_c$ & $(1,1)$ & $1^{+}$ & $7561.5\pm72.5$ & $7752.2\pm75.4$\\
$bc\bar q\bar q$ & $\bar3_c\otimes3_c$ & $(1,1)$ & $2^{+}$ & $7624.5\pm67.3$ & $7815.4\pm69.2$\\
$bq\bar c\bar q$ & $\bar3_c\otimes3_c$ & $(0,0)$ & $0^{+}$ & $7357.2\pm82.3$ & $7561.2\pm84.9$\\
$bq\bar c\bar q$ & $\bar3_c\otimes3_c$ & $(0,1)$ & $1^{+}$ & $7372.8\pm82.5$ & $7584.2\pm85.2$\\
$bq\bar c\bar q$ & $\bar3_c\otimes3_c$ & $(1,0)$ & $1^{+}$ & $7362.6\pm82.3$ & $7569.5\pm85.0$\\
$bq\bar c\bar q$ & $\bar3_c\otimes3_c$ & $(1,1)$ & $0^{+}$ & $7307.6\pm90.5$ & $7523.5\pm93.6$\\
$bq\bar c\bar q$ & $\bar3_c\otimes3_c$ & $(1,1)$ & $1^{+}$ & $7342.9\pm86.4$ & $7557.9\pm89.3$\\
$bq\bar c\bar q$ & $\bar3_c\otimes3_c$ & $(1,1)$ & $2^{+}$ & $7413.5\pm78.8$ & $7626.9\pm81.4$\\
\bottomrule\end{tabular}\end{table*}

% END INLINE tables/ground

For the axial--axial ground multiplet, $bc\bar q\bar q$ gives
$(7530.1,7561.5,7624.5)$ MeV in NR and
$(7720.7,7752.2,7815.4)$ MeV in SR for
$J^P=(0^+,1^+,2^+)$. The corresponding $bq\bar c\bar q$ values are
$(7307.6,7342.9,7413.5)$ and
$(7523.5,7557.9,7626.9)$ MeV. The spin
hierarchy therefore remains stable after propagating the model-parameter errors.

For two axial clusters in an $S$ wave, the diagonal contact interaction
retains the form
\begin{equation}
 M_J=M_{\rm cent}+A d_J,\qquad(d_0,d_1,d_2)=(-2,-1,+1).
 \label{eq:hf}
\end{equation}
The adjacent NR gaps are 31.5 and
63.0 MeV for $bc\bar q\bar q$, and
35.3 and 70.6 MeV for
$bq\bar c\bar q$. The SR gaps are 31.6,
63.2 MeV and 34.5,
69.0 MeV, respectively. Their approximate $1:2$ pattern
is the expected consequence of the contact eigenvalues and is much more
stable than the absolute mass scale.

The additional cluster-spin assignments remain distinct after the full
parameter-error propagation. The isovector scalar--axial $bc\bar q\bar q$ $1^+$ state
is $7587.1\pm69.8$ MeV in NR and
$7775.6\pm72.2$ MeV in SR. For $bq\bar c\bar q$, the scalar--scalar $0^+$ state is
$7357.2\pm82.3$ MeV (NR) and
$7561.2\pm84.9$ MeV (SR). The axial--axial $0^+$ remains the lowest diagonal antitriplet
configuration considered for either flavor sector, but differences among
same-$J^P$ configurations are comparable with the quoted error scale;
they should not be interpreted as precision level separations.

The axial--axial $0^+$ radial sequence also remains nonuniform. For
$bc\bar q\bar q$, the first and second gaps are
581.9 and 390.0 MeV in NR, and
579.1 and 367.3 MeV in SR. For
$bq\bar c\bar q$ they are 594.0,
347.4 MeV (NR) and 605.5,
342.7 MeV (SR). The decreasing second gap is therefore
not an artifact of varying only a restricted subset of parameters.

% BEGIN INLINE tables/spectrum
\begin{table*}[!tbp]
\caption{Antitriplet axial--axial spectra through $3S$, $2P$ and $1D$, in MeV. Entries are quoted as mass $\pm$ propagated parameter-sensitivity error. The complete antitriplet spectra for all cluster-spin configurations appear in Appendix~\ref{app:catalog}.}\label{tab:spectrum}
\centering\scriptsize\setlength{\tabcolsep}{3.2pt}
\begin{tabular}{lcrrrr}\toprule
State & $J^P$ & $bc\bar q\bar q$ NR & $bc\bar q\bar q$ SR & $bq\bar c\bar q$ NR & $bq\bar c\bar q$ SR\\\midrule
$1^{1}S_{0}$ & $0^{+}$ & $7530.1\pm75.4$ & $7720.7\pm78.8$ & $7307.6\pm90.5$ & $7523.5\pm93.6$\\
$1^{3}S_{1}$ & $1^{+}$ & $7561.5\pm72.5$ & $7752.2\pm75.4$ & $7342.9\pm86.4$ & $7557.9\pm89.3$\\
$1^{5}S_{2}$ & $2^{+}$ & $7624.5\pm67.3$ & $7815.4\pm69.2$ & $7413.5\pm78.8$ & $7626.9\pm81.4$\\
$2^{1}S_{0}$ & $0^{+}$ & $8111.9\pm64.2$ & $8299.7\pm63.0$ & $7901.6\pm69.0$ & $8129.0\pm68.3$\\
$2^{3}S_{1}$ & $1^{+}$ & $8126.4\pm63.6$ & $8313.1\pm62.4$ & $7912.7\pm68.5$ & $8139.5\pm67.8$\\
$2^{5}S_{2}$ & $2^{+}$ & $8155.2\pm62.4$ & $8339.8\pm61.1$ & $7934.9\pm67.6$ & $8160.5\pm66.9$\\
$3^{1}S_{0}$ & $0^{+}$ & $8501.9\pm62.8$ & $8667.0\pm60.8$ & $8249.0\pm66.8$ & $8471.7\pm65.4$\\
$3^{3}S_{1}$ & $1^{+}$ & $8512.2\pm62.5$ & $8676.3\pm60.4$ & $8256.2\pm66.5$ & $8478.5\pm65.2$\\
$3^{5}S_{2}$ & $2^{+}$ & $8532.9\pm61.9$ & $8695.0\pm59.8$ & $8270.7\pm66.1$ & $8492.1\pm64.7$\\
$1^{1}P_{1}$ & $1^{-}$ & $7997.0\pm62.7$ & $8200.9\pm61.4$ & $7813.5\pm67.5$ & $8045.6\pm66.8$\\
$1^{3}P_{0}$ & $0^{-}$ & $7969.1\pm64.0$ & $8171.2\pm63.1$ & $7773.5\pm70.2$ & $8005.1\pm69.7$\\
$1^{3}P_{1}$ & $1^{-}$ & $7994.0\pm62.9$ & $8197.4\pm61.7$ & $7808.0\pm67.9$ & $8040.0\pm67.3$\\
$1^{3}P_{2}$ & $2^{-}$ & $8008.8\pm62.2$ & $8213.7\pm60.8$ & $7831.9\pm66.4$ & $8064.3\pm65.6$\\
$1^{5}P_{1}$ & $1^{-}$ & $7978.3\pm63.7$ & $8180.5\pm62.8$ & $7784.4\pm69.6$ & $8016.1\pm69.1$\\
$1^{5}P_{2}$ & $2^{-}$ & $7993.1\pm62.9$ & $8196.8\pm61.7$ & $7808.3\pm67.9$ & $8040.4\pm67.2$\\
$1^{5}P_{3}$ & $3^{-}$ & $8029.8\pm61.6$ & $8236.3\pm60.0$ & $7862.9\pm65.0$ & $8095.8\pm64.0$\\
$2^{1}P_{1}$ & $1^{-}$ & $8394.0\pm61.7$ & $8573.5\pm59.7$ & $8164.4\pm65.8$ & $8391.9\pm64.6$\\
$2^{3}P_{0}$ & $0^{-}$ & $8368.5\pm62.5$ & $8546.8\pm60.8$ & $8130.5\pm67.3$ & $8357.9\pm66.2$\\
$2^{3}P_{1}$ & $1^{-}$ & $8391.4\pm61.9$ & $8570.5\pm59.9$ & $8159.8\pm66.0$ & $8387.3\pm64.8$\\
$2^{3}P_{2}$ & $2^{-}$ & $8405.7\pm61.5$ & $8585.9\pm59.4$ & $8180.4\pm65.2$ & $8408.1\pm63.9$\\
$2^{5}P_{1}$ & $1^{-}$ & $8377.3\pm62.4$ & $8555.6\pm60.5$ & $8140.0\pm67.0$ & $8367.4\pm65.8$\\
$2^{5}P_{2}$ & $2^{-}$ & $8391.7\pm61.9$ & $8571.1\pm59.9$ & $8160.7\pm66.0$ & $8388.2\pm64.8$\\
$2^{5}P_{3}$ & $3^{-}$ & $8426.3\pm61.1$ & $8607.5\pm58.9$ & $8207.5\pm64.5$ & $8435.2\pm63.1$\\
$1^{1}D_{2}$ & $2^{+}$ & $8261.3\pm61.5$ & $8455.8\pm59.6$ & $8061.3\pm65.2$ & $8293.4\pm64.1$\\
$1^{3}D_{1}$ & $1^{+}$ & $8258.0\pm61.7$ & $8451.5\pm59.8$ & $8053.6\pm65.6$ & $8285.6\pm64.5$\\
$1^{3}D_{2}$ & $2^{+}$ & $8262.0\pm61.5$ & $8456.2\pm59.6$ & $8061.0\pm65.3$ & $8293.0\pm64.1$\\
$1^{3}D_{3}$ & $3^{+}$ & $8262.7\pm61.4$ & $8457.8\pm59.4$ & $8065.7\pm65.0$ & $8298.0\pm63.8$\\
$1^{5}D_{0}$ & $0^{+}$ & $8258.6\pm61.8$ & $8451.2\pm60.0$ & $8050.8\pm65.9$ & $8282.6\pm64.8$\\
$1^{5}D_{1}$ & $1^{+}$ & $8256.8\pm61.8$ & $8449.8\pm60.0$ & $8050.1\pm65.8$ & $8282.0\pm64.7$\\
$1^{5}D_{2}$ & $2^{+}$ & $8260.9\pm61.6$ & $8454.5\pm59.7$ & $8057.4\pm65.5$ & $8289.4\pm64.4$\\
$1^{5}D_{3}$ & $3^{+}$ & $8261.5\pm61.5$ & $8456.1\pm59.6$ & $8062.1\pm65.2$ & $8294.3\pm64.0$\\
$1^{5}D_{4}$ & $4^{+}$ & $8266.2\pm61.3$ & $8462.1\pm59.3$ & $8072.8\pm64.7$ & $8305.2\pm63.5$\\
\bottomrule\end{tabular}\end{table*}

% END INLINE tables/spectrum

Orbital excitation generates several states with the same $J^P$ but
different total spin $S$. The table keeps those labels separate because
they form different diagonal basis states and have different recoupling
patterns. Across the complete antitriplet catalog the SR--NR
displacement spans 141.0--206.5 MeV
for $bc\bar q\bar q$ and 203.9--232.8 MeV
for $bq\bar c\bar q$. This scheme dependence remains distinct from the
within-scheme parameter-sensitivity error.

The sextet branch is qualitatively different. With the correct
$k_s=-10/3$ color-singlet coefficient the point-cluster Coulomb attraction
is sufficiently strong that the finite SR momentum expansion is not
controlled for the ground solutions. Quantitative sextet masses,
Regge fits and decay widths are therefore not quoted for this branch
set. This is a limitation of the point-cluster/SR approximation, not a
change of the SU(3) color factor.

\subsection{Placement relative to open-heavy thresholds}
\begin{figure*}[tb]
\includegraphics[width=\textwidth]{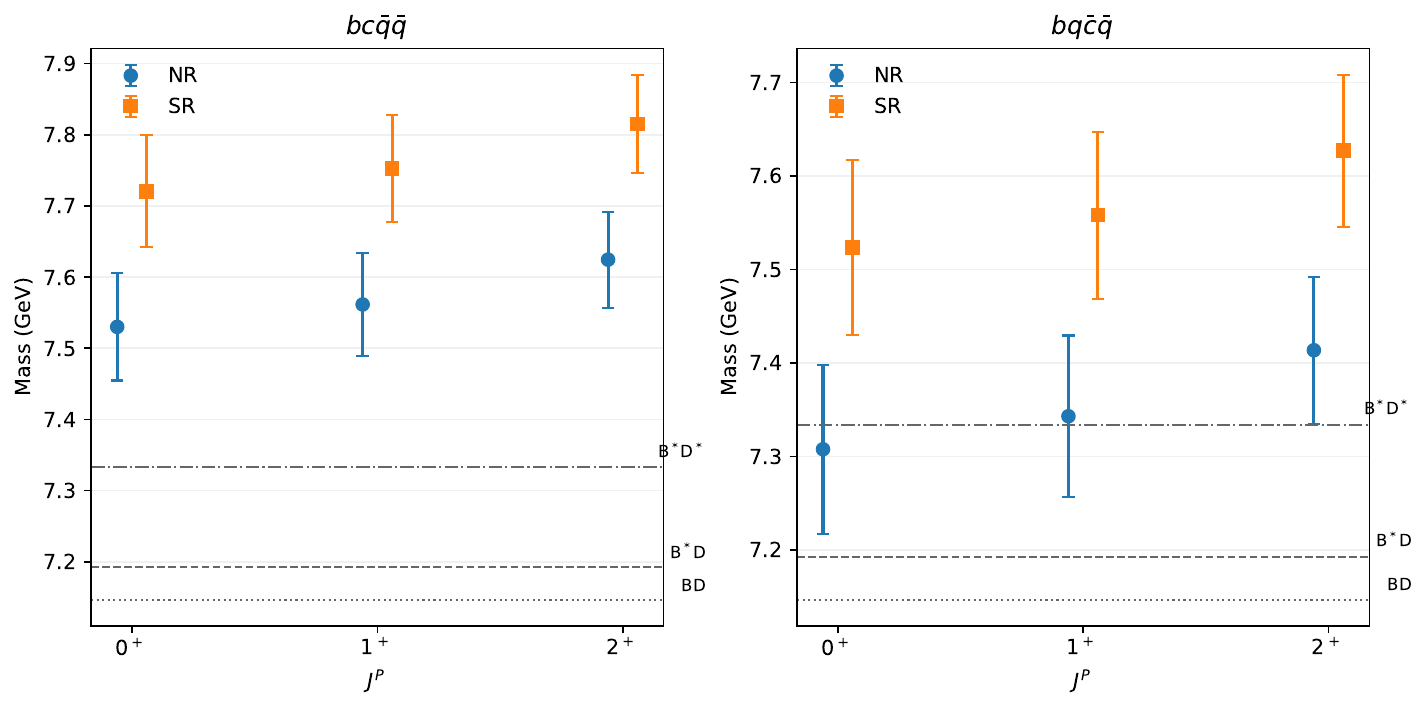}
\caption{Antitriplet axial--axial ground masses with propagated parameter-sensitivity errors. Points show the calculated masses and error bars show the quoted errors. Horizontal lines denote the nominal open-heavy thresholds $BD$, $B^*D$ and $B^*D^*$; the $BD^*$ threshold (7288.1 MeV) lies between the last two lines and is not drawn. The $bc\bar q\bar q$ panel refers to $I=1$; additional $B_c$--light-meson channels for $bq\bar c\bar q$ are given in Table~\ref{tab:decay_bqcq}.}\label{fig:thresholds}\end{figure*}

For each two-meson channel we define
\begin{equation}
 M_{{\rm th},i}=m_{i1}+m_{i2},\qquad
 \Delta_i=M-M_{{\rm th},i}.
 \label{eq:threshold}
\end{equation}
The nominal reference energies are 7146.8 MeV for $BD$,
7192.0 MeV for $B^*D$, 7288.1 MeV for $BD^*$ and
7333.3 MeV for $B^*D^*$. Table~\ref{tab:heavy_decays} uses the calculated parent masses for all
retained ground antitriplet cluster-spin configurations, and
Fig.~\ref{fig:thresholds} displays the axial--axial mass-error bands.
% BEGIN INLINE tables/heavy_decays
\begin{table*}[tb]
\caption{Open-heavy thresholds and rearrangement decays of all retained ground antitriplet configurations using the calculated parent masses. Thresholds $M_{\rm th}$, excess energies $\Delta=M-M_{\rm th}$ and reference widths $\Gamma_i^{(0)}$ are in MeV. Breakup momenta are obtained from Eq.~\eqref{eq:momentum} and are omitted for compactness. Widths follow Eq.~\eqref{eq:referencewidth}; they are reference-normalized visibility estimates, not precision partial widths. A zero width above threshold denotes a vanishing leading $S$-wave spin-recoupling coefficient.}
\label{tab:heavy_decays}
\centering\scriptsize\setlength{\tabcolsep}{3.0pt}
\begin{tabular}{lcclrrrrr}\toprule
System & $(S_d,S_{\bar d})$ & $J^P$ & Channel & $M_{\rm th}$ & $\Delta_{\rm NR}$ & $\Gamma_{\rm NR}^{(0)}$ & $\Delta_{\rm SR}$ & $\Gamma_{\rm SR}^{(0)}$\\\midrule
$bc\bar q\bar q$ & $(0,1)$ & $1^+$ & $B\,D^*$ & 7288.1 & +299.0 & 0.0183 & +487.5 & 0.0225\\
 &  &  & $B^*\,D$ & 7192.0 & +395.1 & 0.0206 & +583.6 & 0.0242\\
 &  &  & $B^*\,D^*$ & 7333.3 & +253.8 & 0.0336 & +442.3 & 0.0427\\
\addlinespace[1pt]
 & $(1,1)$ & $0^+$ & $B\,D$ & 7146.8 & +383.3 & 0.1854 & +573.8 & 0.2186\\
 &  &  & $B^*\,D^*$ & 7333.3 & +196.8 & 0.0449 & +387.4 & 0.0607\\
\addlinespace[1pt]
 & $(1,1)$ & $1^+$ & $B\,D^*$ & 7288.1 & +273.4 & 0.0351 & +464.1 & 0.0441\\
 &  &  & $B^*\,D$ & 7192.0 & +369.5 & 0.0401 & +560.2 & 0.0476\\
 &  &  & $B^*\,D^*$ & 7333.3 & +228.2 & 0.0000 & +418.9 & 0.0000\\
\addlinespace[1pt]
 & $(1,1)$ & $2^+$ & $B^*\,D^*$ & 7333.3 & +291.2 & 0.0429 & +482.1 & 0.0531\\
\addlinespace[1pt]
\midrule
$bq\bar c\bar q$ & $(0,0)$ & $0^+$ & $B\,D$ & 7146.8 & +210.4 & 0.0237 & +414.4 & 0.0319\\
 &  &  & $B^*\,D^*$ & 7333.3 & +23.9 & 0.0243 & +227.9 & 0.0720\\
\addlinespace[1pt]
 & $(0,1)$ & $1^+$ & $B\,D^*$ & 7288.1 & +84.7 & 0.0051 & +296.1 & 0.0091\\
 &  &  & $B^*\,D$ & 7192.0 & +180.8 & 0.0073 & +392.2 & 0.0103\\
 &  &  & $B^*\,D^*$ & 7333.3 & +39.5 & 0.0069 & +250.9 & 0.0167\\
\addlinespace[1pt]
 & $(1,0)$ & $1^+$ & $B\,D^*$ & 7288.1 & +74.5 & 0.0048 & +281.3 & 0.0089\\
 &  &  & $B^*\,D$ & 7192.0 & +170.6 & 0.0071 & +377.5 & 0.0101\\
 &  &  & $B^*\,D^*$ & 7333.3 & +29.3 & 0.0060 & +236.1 & 0.0163\\
\addlinespace[1pt]
 & $(1,1)$ & $0^+$ & $B\,D$ & 7146.8 & +160.8 & 0.0628 & +376.6 & 0.0920\\
 &  &  & $B^*\,D^*$ & 7333.3 & -25.7 & 0.0000 & +190.2 & 0.0221\\
\addlinespace[1pt]
 & $(1,1)$ & $1^+$ & $B\,D^*$ & 7288.1 & +54.8 & 0.0082 & +269.8 & 0.0175\\
 &  &  & $B^*\,D$ & 7192.0 & +150.9 & 0.0134 & +365.9 & 0.0200\\
 &  &  & $B^*\,D^*$ & 7333.3 & +9.6 & 0.0000 & +224.6 & 0.0000\\
\addlinespace[1pt]
 & $(1,1)$ & $2^+$ & $B^*\,D^*$ & 7333.3 & +80.2 & 0.0117 & +293.6 & 0.0215\\
\bottomrule\end{tabular}\end{table*}

% END INLINE tables/heavy_decays

The $bc\bar q\bar q$ axial--axial scalar lies
383.3 MeV above $BD$ and
196.8 MeV above $B^*D^*$ in NR; the SR
excesses are 573.8 and
387.4 MeV. Its leading open-heavy
channels therefore remain open throughout the physically relevant region
of the quoted error band.

For $bq\bar c\bar q$, the NR scalar mass is
25.7 MeV below $B^*D^*$, whereas
the SR mass is 190.2 MeV above it. The
NR separation is only 0.28 times its quoted error, so the sign of this one threshold offset is not a robust model prediction. At the same time, the state remains
160.8 MeV and
376.6 MeV above $BD$ in NR and SR, and
lower $B_c$--light channels are open. Falling below one upper threshold
therefore does not imply stability against strong decay.
 If the $B^*D^*$ reference is instead formed within each scheme from the model
ground-state meson masses of Table~\ref{tab:mesons} (7326.3 MeV in NR and
7341.6 MeV in SR), the axial--axial offsets become $-18.7$ MeV (NR) and
$+181.9$ MeV (SR). The sign pattern is unchanged, but the SR value inherits
the caveat of the daggered $D$ and $B$ entries of Table~\ref{tab:mesons}.

\subsection{Orbital and radial Regge organization}
Regge representations are retained only as compact diagnostics of the
antitriplet excitation pattern. We fit
\begin{equation}M^2=ax+c.\label{eq:regge}\end{equation}
The orbital sequence uses $n=1$, $S=2$ and $x=J=L+2=2,3,4$; radial
sequences use $L=0$, fixed $J=0,1,2$ and $x=n-1=0,1,2$. The fits use the calculated masses.
% BEGIN INLINE tables/regge
\begin{table*}[!tbp]
\caption{Descriptive fits $M^2=ax+c$ to the calculated antitriplet axial--axial masses. $a$, $c$ and $r_{\max}$ are in GeV$^2$; $r_{\max}$ is the maximum absolute residual $|M_{\rm calc}^2-(ax+c)|$ of the three-point fit. Each fit contains three points and is used only as an internal organization diagnostic.}\label{tab:regge}
\centering\small\setlength{\tabcolsep}{4pt}
\begin{tabular}{lllrrrr}\toprule
System & Scheme & Chain & $a$ & $c$ & $R^2$ & $r_{\max}$\\\midrule
$bc\bar q\bar q$ & NR & Orbital, $J=L+2$ & 5.098 & 48.352 & 0.9805 & 0.831\\
$bc\bar q\bar q$ & NR & Radial, $J=0$ & 7.790 & 57.139 & 0.9906 & 0.874\\
$bc\bar q\bar q$ & NR & Radial, $J=1$ & 7.641 & 57.584 & 0.9916 & 0.813\\
$bc\bar q\bar q$ & NR & Radial, $J=2$ & 7.339 & 58.478 & 0.9934 & 0.691\\
$bq\bar c\bar q$ & NR & Orbital, $J=L+2$ & 5.105 & 45.336 & 0.9619 & 1.174\\
$bq\bar c\bar q$ & NR & Radial, $J=0$ & 7.322 & 53.971 & 0.9821 & 1.141\\
$bq\bar c\bar q$ & NR & Radial, $J=1$ & 7.124 & 54.441 & 0.9841 & 1.046\\
$bq\bar c\bar q$ & NR & Radial, $J=2$ & 6.723 & 55.387 & 0.9881 & 0.854\\
$bc\bar q\bar q$ & SR & Orbital, $J=L+2$ & 5.263 & 51.052 & 0.9739 & 0.996\\
$bc\bar q\bar q$ & SR & Radial, $J=0$ & 7.754 & 60.116 & 0.9873 & 1.015\\
$bc\bar q\bar q$ & SR & Radial, $J=1$ & 7.591 & 60.570 & 0.9885 & 0.946\\
$bc\bar q\bar q$ & SR & Radial, $J=2$ & 7.261 & 61.484 & 0.9908 & 0.807\\
$bq\bar c\bar q$ & SR & Orbital, $J=L+2$ & 5.404 & 48.019 & 0.9577 & 1.312\\
$bq\bar c\bar q$ & SR & Radial, $J=0$ & 7.584 & 57.234 & 0.9796 & 1.263\\
$bq\bar c\bar q$ & SR & Radial, $J=1$ & 7.381 & 57.705 & 0.9817 & 1.165\\
$bq\bar c\bar q$ & SR & Radial, $J=2$ & 6.973 & 58.653 & 0.9858 & 0.967\\
\bottomrule\end{tabular}\end{table*}

% END INLINE tables/regge
\begin{figure*}[tb]
\includegraphics[width=\textwidth]{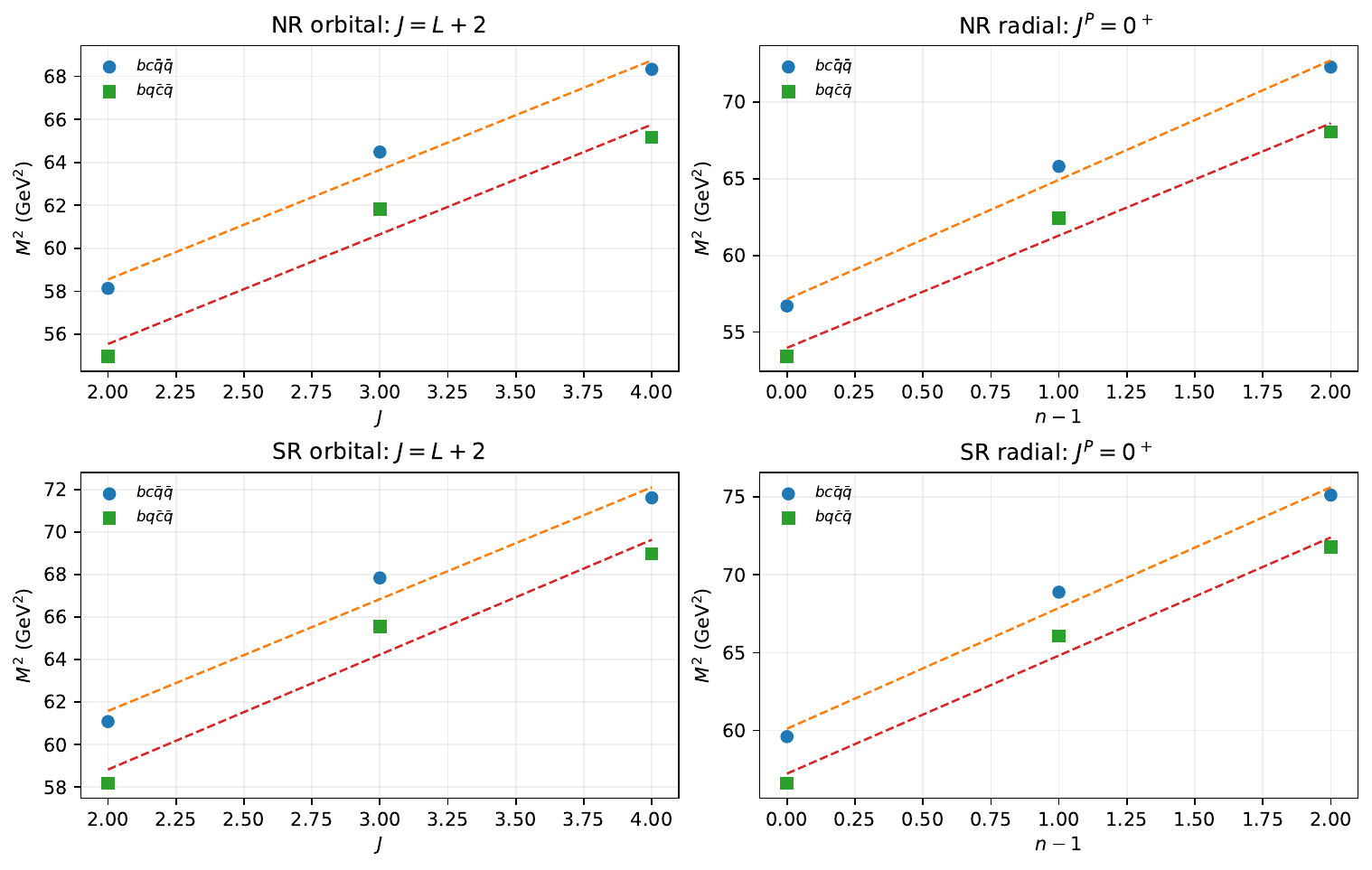}
\caption{Orbital and radial Regge representations of the calculated antitriplet axial--axial masses. Symbols show calculated $M^2$ values (circles: $bc\bar q\bar q$; squares: $bq\bar c\bar q$). Each panel shows a single scheme (NR or SR, as labeled), and the dashed lines are the corresponding least-squares fits for each flavor system (upper line: $bc\bar q\bar q$; lower line: $bq\bar c\bar q$).}
\label{fig:regge}\end{figure*}

The NR orbital slopes are 5.098 and
5.105 GeV$^2$ for $bc\bar q\bar q$ and
$bq\bar c\bar q$, while the SR values are 5.263 and
5.404 GeV$^2$. The radial slopes span
7.339--7.790 GeV$^2$ (NR) and
7.261--7.754 GeV$^2$ (SR) for
$bc\bar q\bar q$, compared with 6.723--7.322
and 6.973--7.584 GeV$^2$ for
$bq\bar c\bar q$. Because every fit contains only three points, the high
$R^2$ values are not evidence for asymptotic linearity; they summarize the
calculated low-lying pattern.
 The calculated $M^2$ values are in fact visibly nonlinear: the successive
increments of the $J=0$ radial chains fall from 9.0--9.5 to 5.6--6.5 GeV$^2$,
and those of the orbital chains from 6.3--7.4 to 3.3--3.9 GeV$^2$ (both
systems, NR and SR). The maximum residual $r_{\max}$ (up to about
1.3 GeV$^2$, corresponding to a mass residual of roughly 80 MeV) is therefore
comparable with the quoted mass errors. The slopes $a$ are much larger than
the $\approx1$ GeV$^2$ of light mesons, as expected for systems with a heavy
cluster, and are not universal.

\section{Rearrangement decays and experimental signatures}\label{sec:threshold}
\subsection{Flavor channels and decay formalism}
Strong rearrangement decays connect the compact configurations to
meson pairs with the same conserved flavor. To make these assignments
explicit, write $B_b=(b\bar q)$, $D_c=(c\bar q)$ and
$D_{\bar c}=(\bar cq)$. The open-heavy channels are
$B_b^{(*)}D_c^{(*)}$ for $\Tbc$ and
$B_b^{(*)}D_{\bar c}^{(*)}$ for $\Tcomp$. The latter also
admits $B_c^-=(b\bar c)$ together with a light $q\bar q$ meson.
In the tables, $B^{(*)}D^{(*)}$ abbreviates the appropriate
open-heavy pair. Charge conjugation gives the same daughter-mass
references for the two sectors, although their flavor assignments
remain distinct.

Table~\ref{tab:daughters} specifies the fixed comparison masses
used in the threshold and phase-space analysis. The conventional
daughter inputs follow the stated PDG-based mass
convention~\cite{PDG}. The $B_c^*$ reference is obtained from
$m_{B_c}=6274.47$ MeV and the lattice hyperfine prediction
$m_{B_c^*}-m_{B_c}=54(3)$ MeV~\cite{Dowdall}, giving
$m_{B_c^*}=6328.47$ MeV. This theoretical vector reference is
separate from the model calibration mass in
Table~\ref{tab:mesons}. The external-mass convention allows the
NR and SR parent estimates to be compared with one common set
of threshold energies.
% BEGIN INLINE tables/daughters
\begin{table}[!tbp]
\caption{Fixed external daughter inputs in MeV. $B,D,D^*$ are charge averages and $\pi$ is the average over its three charge states. The $B_c^*$ value uses $B_c+54(3)$ MeV from lattice QCD; it is not a measured mass. The nominal light-vector masses define phase-space references. See Refs.~\cite{PDG,Dowdall}.}\label{tab:daughters}
\centering\small\setlength{\tabcolsep}{4pt}
\begin{tabular}{lr}\toprule
Daughter & Mass\\\midrule
$B$ & 5279.565\\
$B^*$ & 5324.750\\
$D$ & 1867.250\\
$D^*$ & 2008.555\\
$B_c$ & 6274.470\\
$B_c^*$ & 6328.470\\
$\pi$ & 138.039\\
$\eta$ & 547.862\\
$\eta^\prime$ & 957.780\\
$\rho$ & 775.260\\
$\omega$ & 782.660\\
\bottomrule\end{tabular}\end{table}

% END INLINE tables/daughters

For a parent of mass $M$, the breakup momentum is
\begin{equation}
 p_i(M)=\frac{\sqrt{[M^2-(m_{i1}+m_{i2})^2]
                         [M^2-(m_{i1}-m_{i2})^2]}}{2M},
 \label{eq:momentum}
\end{equation}
when $M>m_{i1}+m_{i2}$. A partial-wave decay amplitude may be
organized into spin, color, flavor and spatial factors,
\begin{equation}
 \mathcal M_i=C_i^{\rm spin}C_i^{\rm color}C_i^{\rm flavor}
                     p_i^{\ell_i}F_i(p_i).
 \label{eq:amplitude}
\end{equation}
With relativistically normalized states, the two-body width has
the corresponding form
\begin{equation}
 \Gamma_i=\frac{p_i}{8\pi M^2}\,
                 \overline{|\mathcal M_i|^2},
 \label{eq:width}
\end{equation}
with the appropriate identical-particle factor where required.
The bar denotes the initial-spin average and final-spin sum.
Equation~\eqref{eq:amplitude} makes the threshold behavior
$\Gamma_i\propto p_i^{2\ell_i+1}$ explicit when the reduced
amplitude is smooth near threshold. A normalization and spatial
transition model are needed to convert this behavior into widths
in MeV.

For the ground positive-parity states, the leading channels have
relative $\ell=0$. Two pseudoscalars then couple to $J=0$,
a pseudoscalar and a vector to $J=1$, and two vectors to
$J=0,1,2$. Here $P$ and $V$ in $PP$, $PV$, $VP$ and $VV$
denote daughter pseudoscalar and vector mesons. A higher partial
wave can allow additional final states: for example, a $2^+$
parent can decay to $PP$ in a relative $D$ wave. The leading
channel registry therefore identifies the lowest orbital barriers
rather than every possible strong process.

\subsection{Spin recoupling and rearrangement widths}
In a spin-independent rearrangement of an $S$-wave parent, the
change from cluster spins to meson spins is determined by
\begin{align}
 C_i^{\rm spin}&=\left\langle[(13)_{s_1}(24)_{s_2}]_J
                \middle|[(12)_{\Sd}(34)_{\Sa}]_J\right\rangle
 \label{eq:recoupling}\\
 &=\widehat S_d\widehat S_{\bar d}\widehat s_1\widehat s_2
 \begin{Bmatrix}
 \tfrac12&\tfrac12&\Sd\\
 \tfrac12&\tfrac12&\Sa\\
 s_1&s_2&J
 \end{Bmatrix} .\nonumber
\end{align}
Here $\widehat j=\sqrt{2j+1}$. The displayed partition gives the $B_c$--light topology, whereas
the open-heavy rearrangement uses $(14)(23)$. For each pure
initial cluster-spin state considered here, the two partitions
give the same squared overlaps. Their relative phases become
relevant if different initial configurations interfere.
% BEGIN INLINE tables/recoupling
\begin{table*}[!tbp]
\caption{Spin and color recoupling for the leading $S$-wave two-meson rearrangement channels. $C_i^{\rm spin}$ is the signed spin amplitude in the $(13)(24)$ phase convention of Eq.~\eqref{eq:recoupling}; the alternative $(14)(23)$ partition gives the same squared overlap for a pure initial configuration. The color-singlet projections are $\mathcal C_{3\text{-}3}^2=1/3$ and $\mathcal C_{6\text{-}6}^2=2/3$. The last two columns give $W_i=|C_i^{\rm spin}|^2\mathcal C_c^2$, before flavor, spatial-overlap and phase-space factors.}\label{tab:recoupling}
\centering\scriptsize\setlength{\tabcolsep}{4.2pt}
\begin{tabular}{llcccc}\toprule
$(S_d,S_{\bar d});J^P$ & Daughter-spin class & $C_i^{\rm spin}$ & $|C_i^{\rm spin}|^2$ & $W_i(\bar3_c\otimes3_c)$ & $W_i(6_c\otimes\bar6_c)$\\\midrule
$(0,0);0^+$ & $PP$ & $+1/2$ & $1/4$ & $1/12$ & $1/6$\\
$(0,0);0^+$ & $VV$ & $+\sqrt3/2$ & $3/4$ & $1/4$ & $1/2$\\
$(0,1);1^+$ & $PV$ & $+1/2$ & $1/4$ & $1/12$ & $1/6$\\
$(0,1);1^+$ & $VP$ & $-1/2$ & $1/4$ & $1/12$ & $1/6$\\
$(0,1);1^+$ & $VV$ & $+1/\sqrt2$ & $1/2$ & $1/6$ & $1/3$\\
$(1,0);1^+$ & $PV$ & $-1/2$ & $1/4$ & $1/12$ & $1/6$\\
$(1,0);1^+$ & $VP$ & $+1/2$ & $1/4$ & $1/12$ & $1/6$\\
$(1,0);1^+$ & $VV$ & $+1/\sqrt2$ & $1/2$ & $1/6$ & $1/3$\\
$(1,1);0^+$ & $PP$ & $+\sqrt3/2$ & $3/4$ & $1/4$ & $1/2$\\
$(1,1);0^+$ & $VV$ & $-1/2$ & $1/4$ & $1/12$ & $1/6$\\
$(1,1);1^+$ & $PV$ & $+1/\sqrt2$ & $1/2$ & $1/6$ & $1/3$\\
$(1,1);1^+$ & $VP$ & $+1/\sqrt2$ & $1/2$ & $1/6$ & $1/3$\\
$(1,1);2^+$ & $VV$ & $1$ & $1$ & $1/3$ & $2/3$\\
\bottomrule\end{tabular}
\end{table*}

% END INLINE tables/recoupling

Table~\ref{tab:recoupling} combines the signed spin amplitudes with the color-dependent channel weights and shows the distinction between spin
assignments with the same $J^P$. The scalar--scalar $J=0$
state has squared $PP:VV$ overlaps $1:3$, whereas the
axial--axial $J=0$ state has the inverse ratio $3:1$.
For $J=1$, either mixed-spin configuration has
$|C_{VV}^{\rm spin}|^2=1/2$, while the pure axial--axial
overlap into two vectors vanishes. The latter is a cancellation
for a specific spin configuration, not a general prohibition
of $1^+\to VV$. The signed axial--axial recouplings and the
corresponding color Fierz decomposition follow the rearrangement
formalism used in our earlier charm and beauty tetraquark studies
\cite{LodhaCharm,LodhaBeauty}; the full four-flavor recoupling is
also discussed in Ref.~\cite{ChengDecay}.

The color part must be recoupled independently of the spin part.
For the two compact singlet bases used here, the earlier decay
formalism gives
\begin{align}
 |\bar3_c3_c\rangle
 &=\sqrt{\frac13}\,|1_c1_c\rangle
  -\sqrt{\frac23}\,|8_c8_c\rangle,\label{eq:colorrec3}\\
 |6_c\bar6_c\rangle
 &=\sqrt{\frac23}\,|1_c1_c\rangle
  +\sqrt{\frac13}\,|8_c8_c\rangle.\label{eq:colorrec6}
\end{align}
Only the $1_c\otimes1_c$ component directly projects onto two
asymptotic color-singlet mesons in the present fall-apart basis.
Consequently,
\begin{equation}
 \mathcal C_{3\text{-}3}^2=\frac13,\qquad
 \mathcal C_{6\text{-}6}^2=\frac23,
 \label{eq:colorsingletprob}
\end{equation}
so the sextet color probability is twice the antitriplet value for
the same spin, flavor, spatial amplitude and parent kinematics.
This factor of two is an amplitude-projection effect only; it does
not imply that a physical sextet width is exactly twice the
antitriplet width, because the parent mass and radial overlap can
also change. The combined spin--color weights for both color sectors are
included directly in Table~\ref{tab:recoupling}.

For numerical decay tables, we specify a reference transition strength
$g_0=1$ GeV and effective couplings $g_J=g_0/\sqrt{2J+1}$.
This $J$ dependence defines the phenomenological normalization used
here; it is not an additional spin average of the normalized overlap
in Eq.~\eqref{eq:recoupling}. The squared color-singlet projections
are $\mathcal C_c^2=1/3$ for $\bar3_c\otimes3_c$ and $2/3$
for $6_c\otimes\bar6_c$. We use the effective topology factor
$\Omega_t^2=1$ for the single open-heavy flavor family of $\Tbc$
and $\Omega_t^2=1/2$ for each of the two distinct flavor families
of $\Tcomp$. The two equivalent light-antiquark pairings in
$\Tbc$ are counted once. These topology factors specify the
reference model and are not measured channel probabilities.

With unit reduced spatial and flavor amplitudes, the reference
$S$-wave widths are
\begin{equation}
 \Gamma_i^{(0)}=\frac{p_i}{8\pi M^2}
 \frac{g_0^2}{2J+1}\,\Omega_t^2\mathcal C_c^2
 |C_i^{\rm spin}|^2\,
 \Theta(M-M_{{\rm th},i}).
 \label{eq:referencewidth}
\end{equation}
All masses and momenta in this expression are evaluated in MeV,
so $g_0=1000$ MeV gives $\Gamma_i^{(0)}$ in MeV. The value
of $g_0$ is a reference normalization, not a fit to tetraquark
decay data. Under the same spin and topology convention, a
different common strength $g$, flavor projection $f_i$ and
spatial amplitude $F_i(p_i)$ give
\begin{equation}
 \Gamma_i=\Gamma_i^{(0)}
 \left|\frac{g}{g_0}\right|^2|f_iF_i(p_i)|^2.
 \label{eq:widthrescale}
\end{equation}
Thus the tabulated numbers are reproducible reference-normalized
model widths. Their absolute scale is not a calibrated physical
prediction. Propagating the parent-mass errors gives conditional width
sensitivities with fixed daughter masses and amplitudes. These are not a
full physical width uncertainty or a total lifetime.

Table~\ref{tab:heavy_decays} gives the leading open-heavy channels for
all retained ground antitriplet cluster-spin configurations. The
$bc\bar q\bar q$ entries include the allowed isovector scalar--axial and
axial--axial states, while the $bq\bar c\bar q$ entries include
scalar--scalar, both mixed-spin assignments and the three axial--axial total-spin
couplings. Table~\ref{tab:decay_bqcq} gives the complementary
$B_c$--light-meson channels for all six $bq\bar c\bar q$ ground states.
Consequently, a given $J^P$ value does not in general specify a unique
internal cluster-spin assignment or decay pattern.

For $bc\bar q\bar q$, the axial--axial scalar reference-normalized $BD$
width is 0.1854 MeV in NR and 0.2186 MeV in SR, compared with
0.0449 and 0.0607 MeV for $B^*D^*$. For the axial--axial
$bq\bar c\bar q$ scalar, the $BD$ entries are 0.0628/0.0920 MeV;
the $B^*D^*$ channel is closed at the NR calculated mass and open in SR.
The complete open-heavy table also makes the configuration dependence
near $B^*D^*$ explicit. In NR, the $bq\bar c\bar q$ scalar--scalar
$0^+$ state lies $23.9$ MeV above this threshold, while the two mixed-spin
$1^+$ states lie $39.5$ and $29.3$ MeV above it; by contrast, the
axial--axial $0^+$ state lies $25.7$ MeV below it. The threshold status is
therefore sensitive to the internal spin assignment as well as to the NR--SR
prescription.

The additional $B_c$--light rows demonstrate the internal-spin
sensitivity directly: the mixed-spin $1^+$ parents carry nonzero
$VV$ strength, whereas the pure axial--axial $1^+$ coefficient
vanishes in the spin-independent recoupling limit. These values are
visibility estimates under the common reference normalization, not
physical branching fractions.
% BEGIN INLINE tables/decay_bqcq
\begin{table*}[!tbp]
\caption{$B_c$--light-meson rearrangement channels of all ground antitriplet $bq\bar c\bar q$ spin configurations using the calculated parent masses. The $I=0$ and $I=1$ rows correspond to distinct light-meson isospins. $\Delta=M-M_{\rm th}$ and the reference-normalized widths are in MeV. Breakup momenta are obtained from Eq.~\eqref{eq:momentum}; they are omitted here to keep all cluster-spin configurations readable in one table. The antitriplet color probability $1/3$ is included through Eq.~\eqref{eq:referencewidth}.}\label{tab:decay_bqcq}
\centering\scriptsize\setlength{\tabcolsep}{4.2pt}
\begin{tabular}{cclcrrrr}\toprule
$(S_d,S_{\bar d})$ & $J^P$ & Channel & $I$ & $\Delta_{\rm NR}$ & $\Gamma_{\rm NR}^{(0)}$ & $\Delta_{\rm SR}$ & $\Gamma_{\rm SR}^{(0)}$\\\midrule
$0,0$ & $0^+$ & $B_c\,\pi$ & 1 & +944.7 & 0.0305 & +1148.7 & 0.0339\\
$0,0$ & $0^+$ & $B_c\,\eta$ & 0 & +534.9 & 0.0265 & +738.9 & 0.0309\\
$0,0$ & $0^+$ & $B_c\,\eta^\prime$ & 0 & +125.0 & 0.0143 & +328.9 & 0.0227\\
$0,0$ & $0^+$ & $B_c^*\,\rho$ & 1 & +253.5 & 0.0577 & +457.5 & 0.0765\\
$0,0$ & $0^+$ & $B_c^*\,\omega$ & 0 & +246.1 & 0.0570 & +450.1 & 0.0760\\
\midrule
$0,1$ & $1^+$ & $B_c\,\rho$ & 1 & +323.1 & 0.0073 & +534.4 & 0.0093\\
$0,1$ & $1^+$ & $B_c\,\omega$ & 0 & +315.7 & 0.0072 & +527.0 & 0.0092\\
$0,1$ & $1^+$ & $B_c^*\,\pi$ & 1 & +906.3 & 0.0098 & +1117.7 & 0.0110\\
$0,1$ & $1^+$ & $B_c^*\,\eta$ & 0 & +496.5 & 0.0084 & +707.8 & 0.0099\\
$0,1$ & $1^+$ & $B_c^*\,\eta^\prime$ & 0 & +86.5 & 0.0039 & +297.9 & 0.0071\\
$0,1$ & $1^+$ & $B_c^*\,\rho$ & 1 & +269.1 & 0.0132 & +480.4 & 0.0174\\
$0,1$ & $1^+$ & $B_c^*\,\omega$ & 0 & +261.7 & 0.0130 & +473.0 & 0.0173\\
\midrule
$1,0$ & $1^+$ & $B_c\,\rho$ & 1 & +312.9 & 0.0072 & +519.7 & 0.0091\\
$1,0$ & $1^+$ & $B_c\,\omega$ & 0 & +305.5 & 0.0071 & +512.3 & 0.0091\\
$1,0$ & $1^+$ & $B_c^*\,\pi$ & 1 & +896.1 & 0.0097 & +1102.9 & 0.0109\\
$1,0$ & $1^+$ & $B_c^*\,\eta$ & 0 & +486.3 & 0.0083 & +693.1 & 0.0099\\
$1,0$ & $1^+$ & $B_c^*\,\eta^\prime$ & 0 & +76.4 & 0.0037 & +283.2 & 0.0070\\
$1,0$ & $1^+$ & $B_c^*\,\rho$ & 1 & +258.9 & 0.0130 & +465.7 & 0.0171\\
$1,0$ & $1^+$ & $B_c^*\,\omega$ & 0 & +251.5 & 0.0128 & +458.3 & 0.0170\\
\midrule
$1,1$ & $0^+$ & $B_c\,\pi$ & 1 & +895.1 & 0.0886 & +1111.0 & 0.1000\\
$1,1$ & $0^+$ & $B_c\,\eta$ & 0 & +485.3 & 0.0757 & +701.1 & 0.0904\\
$1,1$ & $0^+$ & $B_c\,\eta^\prime$ & 0 & +75.3 & 0.0334 & +291.2 & 0.0644\\
$1,1$ & $0^+$ & $B_c^*\,\rho$ & 1 & +203.9 & 0.0173 & +419.7 & 0.0245\\
$1,1$ & $0^+$ & $B_c^*\,\omega$ & 0 & +196.5 & 0.0170 & +412.3 & 0.0243\\
\midrule
$1,1$ & $1^+$ & $B_c\,\rho$ & 1 & +293.2 & 0.0140 & +508.2 & 0.0181\\
$1,1$ & $1^+$ & $B_c\,\omega$ & 0 & +285.8 & 0.0138 & +500.8 & 0.0180\\
$1,1$ & $1^+$ & $B_c^*\,\pi$ & 1 & +876.4 & 0.0192 & +1091.4 & 0.0217\\
$1,1$ & $1^+$ & $B_c^*\,\eta$ & 0 & +466.6 & 0.0163 & +681.6 & 0.0195\\
$1,1$ & $1^+$ & $B_c^*\,\eta^\prime$ & 0 & +56.6 & 0.0064 & +271.7 & 0.0137\\
\midrule
$1,1$ & $2^+$ & $B_c^*\,\rho$ & 1 & +309.8 & 0.0170 & +523.2 & 0.0217\\
$1,1$ & $2^+$ & $B_c^*\,\omega$ & 0 & +302.4 & 0.0168 & +515.8 & 0.0216\\
\bottomrule\end{tabular}
\end{table*}

% END INLINE tables/decay_bqcq

The sextet constants are nevertheless retained explicitly in
Table~\ref{tab:recoupling}. Their two-meson color-singlet probability is
$2/3$, rather than $1/3$, and therefore the combined spin--color
coefficient differs from the antitriplet value for every channel.
Numerical $6_c\otimes\bar6_c$ widths are not quoted because the corrected
$k_s=-10/3$ point-cluster spectrum is not retained as controlled
quantitative spectroscopy in the present analysis.
Quoting a phase-space width from an uncontrolled parent mass would be
less consistent than retaining the exact recoupling coefficient and
withholding the numerical width.

To compare the kinematic patterns without specifying an absolute
transition strength, we define
\begin{align}
 w_i&=|C_i^{\rm spin}|^2\left(\frac{p_i}{p_0}\right)^{2\ell_i+1}
                 \Theta(M-M_{{\rm th},i}),\quad p_0=1\ {\rm GeV},
 \label{eq:weight}\\
 \widehat w_i&=\frac{w_i}{\sum_{j\in\mathcal C}w_j}.
 \label{eq:normal}
\end{align}
The set $\mathcal C$ comprises one stated topology and retained
partial-wave set for a fixed parent. For the leading open-heavy
channels, we use equal reduced spatial amplitudes within that set.
If all channels are closed, no normalized channel weight is assigned.
Common topology-wide color and flavor
factors cancel when at least one channel is open, leaving a comparison of spin recoupling and
available phase space. Channel-dependent spatial overlaps or
coherent exchange amplitudes would modify these ratios.

For the axial--axial $\Tbc$ scalar, $BD$ carries the larger
normalized weight, $0.805$ in NR and $0.783$ in SR, with the
remainder assigned to $B^*D^*$. For the axial parents,
$B^*D$ has more phase space than $BD^*$ because of its lower
threshold. This effect is especially pronounced for $\Tcomp$
in NR, where the normalized $BD^*:B^*D$ weights are
$0.382:0.618$. At the higher SR mass they become
$0.466:0.534$. The comparison illustrates how threshold
placement can change channel balance even when the spin
coefficients are identical.

The weights in Eqs.~\eqref{eq:weight} and \eqref{eq:normal}
are dimensionless. Their purpose is to display the leading
rearrangement pattern under a specified overlap assumption.
They are not absolute widths or measured branching fractions.
The spatial normalization for the present heavy-flavor transitions must
be determined in its own dynamical calculation.

\subsection{Isospin-resolved channels and experimental signatures}
The $B_c$--light-meson channels distinguish the light isospin
of $\Tcomp$. The $\pi$ and $\rho$ channels belong to $I=1$,
while $\eta$, $\eta'$ and $\omega$ belong to $I=0$.
Table~\ref{tab:decay_bqcq} lists the ground-state channels for all
antitriplet cluster-spin assignments, together with their excess
energies and reference widths. Channels assigned to different
parent isospins are alternatives and are not combined into one total
width. The scalar--scalar and axial--axial $0^+$ parents should in
particular not be conflated: the former has $PP:VV=1:3$ in spin
probability, while the latter has $3:1$.

For the axial--axial scalar, the $B_c\pi$ excesses are
895.1 and 1111.0 MeV in NR and SR, while the $B_c\eta$ excesses are
485.3 and 701.1 MeV. By comparison, the $B_c\eta'$ excess increases from
75.3 to 291.2 MeV, and the axial--axial $B_c^*\eta'$ excess changes from
56.6 to 271.7 MeV. The mixed-spin $1^+$ rows add $B_c^*\rho$ and
$B_c^*\omega$ contributions that are absent for a pure axial--axial
$1^+$ parent. These channels remain especially sensitive to the
parent-mass prescription. For a nonstrange isoscalar
component $\eta_q=(u\bar u+d\bar d)/\sqrt2$, a chosen
$\eta$--$\eta'$ mixing convention would add the associated
$\cos^2\phi$ and $\sin^2\phi$ flavor factors. The reference
widths use $f_i=1$; those physical flavor projections can be
applied through Eq.~\eqref{eq:widthrescale}. No mixing angle
is assumed in the numerical columns.

Broad vector daughters require an additional qualification.
For example, the nominal $B_c\rho$ mass sum is a convenient
reference energy, but the observed channel contains the
corresponding multipion spectrum. A quantitative threshold
shape must fold the transition amplitude with the daughter
spectral function. The narrow-daughter expression in
Eq.~\eqref{eq:momentum} is used here to compare nominal
two-body phase space.

Experimentally, charge-related $B_c\pi$ modes offer a way
to test an isovector assignment, while $B_c\eta$ and
$B_c\eta'$ provide complementary isoscalar channels.
Open-heavy $BD$ and $B^{(*)}D^{(*)}$ combinations probe the
same mass region through a different rearrangement topology.
Vector daughters may involve soft photons or pions in their
decay chains; these must be included when reconstructing
the parent mass and comparing channel yields. Conventional
$B_c$ spectroscopy and decay studies supply the relevant
daughter-system context~\cite{PatelBc}. The most useful
comparison would combine invariant-mass and angular
information in several flavor- and isospin-compatible final
states, with efficiencies and production rates determined
separately from the compact spectrum.

\section{Results and discussion}\label{sec:discussion}
\subsection{The \texorpdfstring{$bc\bar q\bar q$}{bcqq} system}
The six-parameter sensitivity analysis changes the numerical interpretation
more than the central spectroscopy pattern. The antitriplet isovector
axial--axial scalar is $7530.1\pm75.4$ MeV in NR and
$7720.7\pm78.8$ MeV in SR, while the $1^+$ and $2^+$ partners occur at
$7561.5\pm72.5$ and $7624.5\pm67.3$ MeV in NR and at
$7752.2\pm75.4$ and $7815.4\pm69.2$ MeV in SR. The successive central
spin gaps are therefore about 31 and 63 MeV in both prescriptions. These
splittings are smaller than the quoted absolute-mass errors, so the
individual intervals overlap; the physically useful result is the systematic
$0^+<1^+<2^+$ ordering generated by the common spin interaction rather than
three statistically isolated mass windows.

The excitation pattern is much more widely separated. For the axial--axial
$0^+$ chain, the NR radial gaps are about 582 and 390 MeV and the SR gaps
about 579 and 367 MeV. The first $P$- and $D$-wave shells cluster near
8.0--8.3 GeV in NR and 8.2--8.5 GeV in SR. Thus confinement controls the
large shell spacing while spin terms generate the finer structure within
each shell; the Regge fits summarize this low-lying organization rather than
provide an asymptotic extrapolation.

All three ground axial--axial levels remain well above their leading
open-heavy thresholds. For the scalar, the central NR mass is about 383 MeV
above $BD$ and 197 MeV above $B^*D^*$; in SR the offsets increase to about
574 and 387 MeV. These states are therefore compact basis states embedded in
open meson channels rather than candidates for strong stability. Their decay
patterns nevertheless retain structural information: the scalar couples to
$BD$ and $B^*D^*$, the axial state to $B^*D$ and $BD^*$, and the tensor to
spin-two $B^*D^*$. The mixed $(0,1)$ $1^+$ state also has a nonzero leading
$VV$ overlap, so the axial--axial $1^+\to VV$ zero is a diagnostic of a pure
spin configuration rather than a universal $1^+$ selection rule.

The reference-normalized widths should therefore be compared as relative
channel patterns rather than absolute partial widths. Correlated signals in
several $B^{(*)}D^{(*)}$ combinations would test the spin assignment more
strongly than a mass match alone, because the recoupling zeros and relative
weights follow directly from the basis algebra.

Comparison with lattice QCD also requires an explicit isospin match. The
present $bc\bar q\bar q$ calculation uses the isovector light-antidiquark
configuration, while the low-lying lattice states in
Refs.~\cite{RadhakrishnanScalar,Padmanath} are isoscalar. The positive
isovector threshold offsets obtained here therefore neither reproduce nor
exclude those isoscalar poles. Lattice and chiral-EFT studies likewise point
to different dynamics in the $I=0$ and $I=1$ heavy-meson channels
\cite{Abolnikov,LiuChEFT}. A natural extension is therefore to introduce the
isoscalar scalar light-antidiquark basis and couple it explicitly to the
corresponding $B^{(*)}D^{(*)}$ continuum.

The sextet coefficient remains $k_s=-10/3$, as required by SU(3), but the
strong attraction drives the point-cluster SR expansion outside its
controlled domain. We therefore do not interpret sextet masses or widths as
quantitative predictions. The exact rearrangement algebra is still useful:
the $6_c\otimes\bar6_c$ two-meson singlet probability is $2/3$, compared
with $1/3$ for $\bar3_c\otimes3_c$, and Table~\ref{tab:recoupling} shows how
this combines with the spin overlaps. A finite-size, fully relativistic or
common four-body color-basis treatment is needed before the sextet component
can be connected reliably to physical poles.

\subsection{The \texorpdfstring{$bq\bar c\bar q$}{bqcq} system}
The antitriplet $bq\bar c\bar q$ spectrum is lower than the corresponding
$bc\bar q\bar q$ spectrum in the present clustering scheme. For the
axial--axial scalar the difference is approximately 222.5 MeV in NR and
197.2 MeV in SR. This comparison concerns distinct conserved-flavor sectors,
not a mixing matrix element. Within $bq\bar c\bar q$, however,
scalar--scalar, mixed-spin and axial--axial configurations can share the same
$J^P$ and should ultimately be coupled.

The ground basis is already dense. In NR the axial--axial $0^+$ state is
$7307.6\pm90.5$ MeV, the scalar--scalar $0^+$ state is near 7357 MeV, and
the two mixed-spin $1^+$ states lie near 7363 and 7373 MeV. Their separations
are only a few tens of MeV, smaller than the absolute parameter-sensitivity
errors. A physical state in this region therefore need not coincide with one
basis row; spin, color and continuum mixing can redistribute both masses and
decay strength.

The much larger radial gaps, roughly 0.35--0.61 GeV, show that confinement
still dominates the shell spacing while spin interactions resolve the nearby
ground levels.

The $B^*D^*$ threshold is particularly important because several NR ground
configurations lie close to it. The axial--axial $0^+$ state is 25.7 MeV
below threshold, the scalar--scalar $0^+$ is about 23.9 MeV above, the
mixed-spin $1^+$ states are about 29--40 MeV above, and the axial--axial
$1^+$ only about 10 MeV above. These offsets are small compared with the
quoted errors. In SR the same configurations lie clearly above $B^*D^*$,
mostly by 0.19--0.30 GeV. The NR result therefore identifies a
near-threshold multiplet for which coupled-channel dynamics may be important,
rather than a single isolated candidate.

Lower $BD$ and related channels remain accessible, so the NR scalar is not
strongly stable merely because its central value falls below $B^*D^*$.
Nearby continua can shift poles and redistribute widths, which is why compact
model eigenvalues must be compared cautiously with lattice finite-volume
levels and scattering poles~\cite{Alexandrou,Padmanath,Abolnikov}.

The additional $B_c$--light topology gives complementary isospin and spin
information. Isovector states can appear in $B_c\pi$, whereas isoscalar
partners can feed $B_c\eta$ and $B_c\eta'$. The $\eta'$ modes are among the
closest to threshold: the axial--axial $0^+$ state has an NR $B_c\eta'$
excess of about 75 MeV and the axial--axial $1^+$ $B_c^*\eta'$ excess is
about 57 MeV. Moreover, scalar--scalar and axial--axial $0^+$ states have
inverse $PP:VV$ overlap hierarchies, while mixed-spin $1^+$ states permit a
$VV$ component absent for a pure axial--axial $1^+$. These correlated
channel patterns can help distinguish states whose masses alone overlap.

The calculation remains a compact-model sensitivity analysis rather than a
pole determination. The quoted errors vary six parameters of the fixed
Hamiltonian but do not include changes of potential form, finite cluster
size, color mixing or continuum self-energies. The NR--SR difference is a
separate scheme dependence. Because several spin configurations and
thresholds are simultaneously nearby, $bq\bar c\bar q$ is particularly
well suited for a coupled-channel calculation of masses, line shapes and
branching patterns from a common amplitude.

Several further limitations should be kept in mind. (i) The fixed parameter
sets are calibrated only on conventional $D$, $B$ and $B_c$ states; no
tetraquark mass is used as an anchor, so the absolute scale is untested.
(ii) The NR and SR diquark confinement strengths are set to
$b_{\rm meson}/4$ and $b_{\rm meson}/2$, respectively, and the SR diquark
masses of Table~\ref{tab:diquarks} lie 106--167 MeV above the NR ones; the
SR--NR displacement of the tetraquark masses therefore reflects the
diquark-stage inputs as well as the kinetic prescription. (iii) The
intercluster spin--spin and spin--orbit terms are written in terms of cluster
spins and cluster masses [Eqs.~\eqref{eq:ss} and \eqref{eq:ls}]; a
constituent-level treatment would give different effective couplings,
particularly for a light antidiquark, and the contact term vanishes by
construction when either cluster is scalar, so the near-degeneracy of the
scalar--scalar and mixed-spin ground rows in $bq\bar c\bar q$ is partly
structural. (iv) For $bq\bar c\bar q$ with light quarks of equal flavor,
annihilation and mixing with conventional excited $B_c$ states, which may lie
in the same mass region, are not included.

\section{Summary and conclusions}\label{sec:conclusion}
We have studied the $bc\bar q\bar q$ and $bq\bar c\bar q$ spectra in a
Cornell diquark--antidiquark framework with NR and SR kinematics. Errors from
$m_1$, $m_2$, $\alpha_s$, $\sigma$, $b$ and $V_0$ are propagated through the
radial, spin-dependent and kinetic terms before the masses are formed. The
reported $\Delta M$ values are therefore state-dependent parameter-sensitivity
errors rather than a common error assigned to spin partners.

The antitriplet axial--axial ground multiplets retain
$0^+<1^+<2^+$. The $bc\bar q\bar q$ scalar is
$7530.1\pm75.4$ MeV (NR) and $7720.7\pm78.8$ MeV (SR), while the
$bq\bar c\bar q$ scalar is $7307.6\pm90.5$ and
$7523.5\pm93.6$ MeV. Radial gaps decrease with excitation, and the Regge
fits continue to provide a useful low-lying organizational diagnostic.

The most threshold-sensitive result is the NR $bq\bar c\bar q$ scalar:
its calculated mass lies about 25.7 MeV below $B^*D^*$, substantially less
than its 90.5 MeV quoted error. Its SR mass lies about 190.2 MeV above that
threshold. Lower strong channels remain open in both cases, and
reference-normalized rearrangement widths should be read only as
channel-visibility estimates.

Spin and color recoupling enter as separate factors. The sextet two-meson
singlet probability is $2/3$, compared with $1/3$ for the antitriplet, while
the spin amplitudes depend on $(S_d,S_{\bar d},J)$. Their combination fixes
the channel-dependent spin--color weights used in the rearrangement analysis.

Finally, the sextet coefficient is kept at the correct SU(3) value
$k_s=-10/3$. Because this attraction pushes the finite SR expansion outside
its controlled range in the point-cluster treatment, sextet spectroscopy and
numerical widths are not promoted to quantitative predictions. The exact
spin--color recoupling coefficients are nevertheless retained. The next
physical steps are a finite-size or fully relativistic treatment, color
mixing and explicit coupling to the open two-hadron channels.

On the experimental side, open-flavor bottom--charm tetraquarks require an
environment with substantial simultaneous beauty and charm production. The
HL-LHC program, and especially LHCb Upgrade II, is therefore a natural
near-term setting: its projected high-luminosity data set and dedicated
heavy-flavor acceptance are expected to extend hadron-spectroscopy and
exotic-hadron searches substantially~\cite{LHCbUpgradeII}. At a later stage,
future $e^+e^-$ $Z$ factories could offer a complementary low-background
environment with very large samples of both beauty and charm hadrons.
Heavy-flavor studies at FCC-ee have been identified as a major opportunity
of the proposed Tera-$Z$ program~\cite{FCCeeHeavy}, while dedicated CEPC
studies already demonstrate the potential of a high-statistics $Z$ factory
for producing and reconstructing heavy exotic systems~\cite{CEPCExotics}.
For the states considered here, searches should emphasize correlated
$B^{(*)}D^{(*)}$ and $B_c$--light-meson invariant masses together with
angular information and isospin-related channels, since the predicted
spin--color recoupling patterns provide experimentally testable relations
beyond a single mass peak.

\section*{AI Use Statement}
Anthropic's Claude (Claude Sonnet 5.5) was used only to improve the coherence
and readability of the manuscript. The authors take full responsibility for
the content of the manuscript.

\section*{Data Availability Statement}
The numerical data and custom computational material required to support and
reproduce the tables and figures are not deposited in a public repository at
present because they form part of an ongoing computational research program.
They are available from the authors upon reasonable request.

\clearpage
\appendix
\section{Complete mass spectra}\label{app:catalog}
Tables~\ref{tab:fullbcqq} and \ref{tab:fullbqcq} give the complete
antitriplet NR and SR spectra for $n=1,2,3$ in the $S$, $P$ and $D$
sectors. Each entry is quoted as $M\pm\Delta M$, rounded to whole MeV for readability; $\Delta M$ denotes the propagated parameter-sensitivity error. The $bc\bar q\bar q$
entries refer to $I=1$. The sextet branch is excluded from the quantitative
catalog for the convergence reason discussed in Sec.~\ref{sec:data}.

{\small\setlength{\tabcolsep}{3pt}
% BEGIN INLINE tables/full_bcqq
\begin{longtable}{ccrrrrrr}
\caption{Complete $bc\bar q\bar q$ antitriplet mass spectrum. Masses and propagated errors are in MeV; each row contains three radial levels in both prescriptions. $S$, $P$ and $D$ waves have parity $+$, $-$ and $+$, respectively. The sextet branch is not included in this quantitative catalog because the correct $k_s=-10/3$ short-distance attraction makes the finite SR expansion uncontrolled.}\label{tab:fullbcqq}\\
\toprule $(S_d,S_{\bar d})$ & ${}^{2S+1}L_J$ & $1\mathrm{NR}$ & $1\mathrm{SR}$ & $2\mathrm{NR}$ & $2\mathrm{SR}$ & $3\mathrm{NR}$ & $3\mathrm{SR}$\\\midrule\endfirsthead
\multicolumn{8}{c}{\tablename\ \thetable\ continued}\\\toprule $(S_d,S_{\bar d})$ & ${}^{2S+1}L_J$ & $1\mathrm{NR}$ & $1\mathrm{SR}$ & $2\mathrm{NR}$ & $2\mathrm{SR}$ & $3\mathrm{NR}$ & $3\mathrm{SR}$\\\midrule\endhead
\midrule\multicolumn{8}{r}{Continued on next page}\\\endfoot
\bottomrule\endlastfoot
\multicolumn{8}{c}{$\bar3_c\otimes3_c$}\\\midrule
\multicolumn{8}{c}{$ (S_d,S_{\bar d})=(0,1) $}\\\midrule
$(0,1)$ & $^{3}S_{1}$ & $7587\pm70$ & $7776\pm72$ & $8135\pm63$ & $8318\pm62$ & $8517\pm62$ & $8677\pm60$\\
$(0,1)$ & $^{3}P_{0}$ & $7975\pm63$ & $8176\pm62$ & $8374\pm62$ & $8550\pm60$ & $8705\pm62$ & $8857\pm60$\\
$(0,1)$ & $^{3}P_{1}$ & $7986\pm63$ & $8187\pm62$ & $8384\pm62$ & $8561\pm60$ & $8715\pm62$ & $8867\pm60$\\
$(0,1)$ & $^{3}P_{2}$ & $8006\pm62$ & $8209\pm61$ & $8403\pm61$ & $8582\pm59$ & $8734\pm62$ & $8887\pm59$\\
$(0,1)$ & $^{3}D_{1}$ & $8253\pm62$ & $8445\pm60$ & $8596\pm62$ & $8761\pm59$ & $8897\pm62$ & $9038\pm60$\\
$(0,1)$ & $^{3}D_{2}$ & $8255\pm62$ & $8447\pm60$ & $8598\pm62$ & $8764\pm59$ & $8900\pm62$ & $9042\pm60$\\
$(0,1)$ & $^{3}D_{3}$ & $8257\pm61$ & $8450\pm59$ & $8602\pm61$ & $8768\pm59$ & $8904\pm62$ & $9046\pm60$\\
\addlinespace[3pt]
\multicolumn{8}{c}{$ (S_d,S_{\bar d})=(1,1) $}\\\midrule
$(1,1)$ & $^{1}S_{0}$ & $7530\pm75$ & $7721\pm79$ & $8112\pm64$ & $8300\pm63$ & $8502\pm63$ & $8667\pm61$\\
$(1,1)$ & $^{3}S_{1}$ & $7562\pm72$ & $7752\pm75$ & $8126\pm64$ & $8313\pm62$ & $8512\pm62$ & $8676\pm60$\\
$(1,1)$ & $^{5}S_{2}$ & $7625\pm67$ & $7815\pm69$ & $8155\pm62$ & $8340\pm61$ & $8533\pm62$ & $8695\pm60$\\
$(1,1)$ & $^{1}P_{1}$ & $7997\pm63$ & $8201\pm61$ & $8394\pm62$ & $8574\pm60$ & $8725\pm62$ & $8879\pm60$\\
$(1,1)$ & $^{3}P_{0}$ & $7969\pm64$ & $8171\pm63$ & $8369\pm63$ & $8547\pm61$ & $8700\pm63$ & $8854\pm60$\\
$(1,1)$ & $^{3}P_{1}$ & $7994\pm63$ & $8197\pm62$ & $8391\pm62$ & $8571\pm60$ & $8722\pm62$ & $8876\pm60$\\
$(1,1)$ & $^{3}P_{2}$ & $8009\pm62$ & $8214\pm61$ & $8406\pm61$ & $8586\pm59$ & $8736\pm62$ & $8891\pm59$\\
$(1,1)$ & $^{5}P_{1}$ & $7978\pm64$ & $8181\pm63$ & $8377\pm62$ & $8556\pm61$ & $8709\pm62$ & $8862\pm60$\\
$(1,1)$ & $^{5}P_{2}$ & $7993\pm63$ & $8197\pm62$ & $8392\pm62$ & $8571\pm60$ & $8723\pm62$ & $8877\pm60$\\
$(1,1)$ & $^{5}P_{3}$ & $8030\pm62$ & $8236\pm60$ & $8426\pm61$ & $8607\pm59$ & $8756\pm62$ & $8912\pm59$\\
$(1,1)$ & $^{1}D_{2}$ & $8261\pm61$ & $8456\pm60$ & $8605\pm62$ & $8773\pm59$ & $8907\pm62$ & $9050\pm60$\\
$(1,1)$ & $^{3}D_{1}$ & $8258\pm62$ & $8451\pm60$ & $8600\pm62$ & $8768\pm59$ & $8902\pm62$ & $9045\pm60$\\
$(1,1)$ & $^{3}D_{2}$ & $8262\pm61$ & $8456\pm60$ & $8605\pm62$ & $8773\pm59$ & $8907\pm62$ & $9050\pm60$\\
$(1,1)$ & $^{3}D_{3}$ & $8263\pm61$ & $8458\pm59$ & $8607\pm61$ & $8776\pm59$ & $8909\pm62$ & $9054\pm60$\\
$(1,1)$ & $^{5}D_{0}$ & $8259\pm62$ & $8451\pm60$ & $8600\pm62$ & $8767\pm59$ & $8901\pm63$ & $9044\pm60$\\
$(1,1)$ & $^{5}D_{1}$ & $8257\pm62$ & $8450\pm60$ & $8599\pm62$ & $8766\pm59$ & $8900\pm63$ & $9043\pm60$\\
$(1,1)$ & $^{5}D_{2}$ & $8261\pm62$ & $8454\pm60$ & $8603\pm62$ & $8771\pm59$ & $8905\pm62$ & $9048\pm60$\\
$(1,1)$ & $^{5}D_{3}$ & $8262\pm61$ & $8456\pm60$ & $8605\pm62$ & $8774\pm59$ & $8908\pm62$ & $9052\pm60$\\
$(1,1)$ & $^{5}D_{4}$ & $8266\pm61$ & $8462\pm59$ & $8611\pm61$ & $8781\pm59$ & $8914\pm62$ & $9059\pm59$\\
\addlinespace[3pt]
\end{longtable}

% END INLINE tables/full_bcqq
\clearpage
% BEGIN INLINE tables/full_bqcq
\begin{longtable}{ccrrrrrr}
\caption{Complete $bq\bar c\bar q$ antitriplet mass spectrum. Masses and propagated errors are in MeV; each row contains three radial levels in both prescriptions. $S$, $P$ and $D$ waves have parity $+$, $-$ and $+$, respectively. The sextet branch is not included in this quantitative catalog because the correct $k_s=-10/3$ short-distance attraction makes the finite SR expansion uncontrolled.}\label{tab:fullbqcq}\\
\toprule $(S_d,S_{\bar d})$ & ${}^{2S+1}L_J$ & $1\mathrm{NR}$ & $1\mathrm{SR}$ & $2\mathrm{NR}$ & $2\mathrm{SR}$ & $3\mathrm{NR}$ & $3\mathrm{SR}$\\\midrule\endfirsthead
\multicolumn{8}{c}{\tablename\ \thetable\ continued}\\\toprule $(S_d,S_{\bar d})$ & ${}^{2S+1}L_J$ & $1\mathrm{NR}$ & $1\mathrm{SR}$ & $2\mathrm{NR}$ & $2\mathrm{SR}$ & $3\mathrm{NR}$ & $3\mathrm{SR}$\\\midrule\endhead
\midrule\multicolumn{8}{r}{Continued on next page}\\\endfoot
\bottomrule\endlastfoot
\multicolumn{8}{c}{$\bar3_c\otimes3_c$}\\\midrule
\multicolumn{8}{c}{$ (S_d,S_{\bar d})=(0,0) $}\\\midrule
$(0,0)$ & $^{1}S_{0}$ & $7357\pm82$ & $7561\pm85$ & $7903\pm68$ & $8118\pm67$ & $8243\pm66$ & $8454\pm65$\\
$(0,0)$ & $^{1}P_{1}$ & $7800\pm67$ & $8021\pm66$ & $8150\pm66$ & $8367\pm64$ & $8435\pm65$ & $8645\pm64$\\
$(0,0)$ & $^{1}D_{2}$ & $8041\pm65$ & $8262\pm64$ & $8335\pm65$ & $8550\pm63$ & $8591\pm65$ & $8799\pm64$\\
\addlinespace[3pt]
\multicolumn{8}{c}{$ (S_d,S_{\bar d})=(0,1) $}\\\midrule
$(0,1)$ & $^{3}S_{1}$ & $7373\pm82$ & $7584\pm85$ & $7918\pm68$ & $8142\pm67$ & $8258\pm66$ & $8477\pm65$\\
$(0,1)$ & $^{3}P_{0}$ & $7784\pm69$ & $8013\pm69$ & $8139\pm67$ & $8364\pm66$ & $8425\pm66$ & $8643\pm65$\\
$(0,1)$ & $^{3}P_{1}$ & $7800\pm68$ & $8029\pm67$ & $8153\pm66$ & $8377\pm65$ & $8438\pm66$ & $8655\pm64$\\
$(0,1)$ & $^{3}P_{2}$ & $7832\pm66$ & $8061\pm65$ & $8180\pm65$ & $8404\pm64$ & $8463\pm65$ & $8681\pm63$\\
$(0,1)$ & $^{3}D_{1}$ & $8050\pm66$ & $8279\pm64$ & $8344\pm65$ & $8566\pm64$ & $8599\pm66$ & $8814\pm64$\\
$(0,1)$ & $^{3}D_{2}$ & $8055\pm65$ & $8284\pm64$ & $8348\pm65$ & $8571\pm64$ & $8604\pm66$ & $8819\pm64$\\
$(0,1)$ & $^{3}D_{3}$ & $8061\pm65$ & $8291\pm64$ & $8355\pm65$ & $8578\pm63$ & $8611\pm65$ & $8826\pm64$\\
\addlinespace[3pt]
\multicolumn{8}{c}{$ (S_d,S_{\bar d})=(1,0) $}\\\midrule
$(1,0)$ & $^{3}S_{1}$ & $7363\pm82$ & $7569\pm85$ & $7908\pm68$ & $8126\pm67$ & $8248\pm66$ & $8462\pm65$\\
$(1,0)$ & $^{3}P_{0}$ & $7774\pm69$ & $7998\pm68$ & $8129\pm67$ & $8349\pm65$ & $8415\pm66$ & $8628\pm65$\\
$(1,0)$ & $^{3}P_{1}$ & $7789\pm68$ & $8014\pm67$ & $8142\pm66$ & $8362\pm65$ & $8428\pm66$ & $8641\pm64$\\
$(1,0)$ & $^{3}P_{2}$ & $7821\pm66$ & $8045\pm65$ & $8169\pm65$ & $8389\pm64$ & $8453\pm65$ & $8666\pm63$\\
$(1,0)$ & $^{3}D_{1}$ & $8039\pm66$ & $8264\pm64$ & $8334\pm65$ & $8551\pm64$ & $8589\pm66$ & $8800\pm64$\\
$(1,0)$ & $^{3}D_{2}$ & $8044\pm65$ & $8268\pm64$ & $8338\pm65$ & $8556\pm64$ & $8594\pm66$ & $8804\pm64$\\
$(1,0)$ & $^{3}D_{3}$ & $8050\pm65$ & $8275\pm64$ & $8345\pm65$ & $8563\pm63$ & $8601\pm65$ & $8812\pm63$\\
\addlinespace[3pt]
\multicolumn{8}{c}{$ (S_d,S_{\bar d})=(1,1) $}\\\midrule
$(1,1)$ & $^{1}S_{0}$ & $7308\pm91$ & $7523\pm94$ & $7902\pm69$ & $8129\pm68$ & $8249\pm67$ & $8472\pm65$\\
$(1,1)$ & $^{3}S_{1}$ & $7343\pm86$ & $7558\pm89$ & $7913\pm69$ & $8139\pm68$ & $8256\pm67$ & $8479\pm65$\\
$(1,1)$ & $^{5}S_{2}$ & $7413\pm79$ & $7627\pm81$ & $7935\pm68$ & $8160\pm67$ & $8271\pm66$ & $8492\pm65$\\
$(1,1)$ & $^{1}P_{1}$ & $7814\pm67$ & $8046\pm67$ & $8164\pm66$ & $8392\pm65$ & $8449\pm66$ & $8670\pm64$\\
$(1,1)$ & $^{3}P_{0}$ & $7773\pm70$ & $8005\pm70$ & $8130\pm67$ & $8358\pm66$ & $8417\pm67$ & $8638\pm65$\\
$(1,1)$ & $^{3}P_{1}$ & $7808\pm68$ & $8040\pm67$ & $8160\pm66$ & $8387\pm65$ & $8444\pm66$ & $8665\pm64$\\
$(1,1)$ & $^{3}P_{2}$ & $7832\pm66$ & $8064\pm66$ & $8180\pm65$ & $8408\pm64$ & $8464\pm65$ & $8685\pm64$\\
$(1,1)$ & $^{5}P_{1}$ & $7784\pm70$ & $8016\pm69$ & $8140\pm67$ & $8367\pm66$ & $8426\pm66$ & $8647\pm65$\\
$(1,1)$ & $^{5}P_{2}$ & $7808\pm68$ & $8040\pm67$ & $8161\pm66$ & $8388\pm65$ & $8445\pm66$ & $8666\pm64$\\
$(1,1)$ & $^{5}P_{3}$ & $7863\pm65$ & $8096\pm64$ & $8207\pm64$ & $8435\pm63$ & $8489\pm65$ & $8710\pm63$\\
$(1,1)$ & $^{1}D_{2}$ & $8061\pm65$ & $8293\pm64$ & $8355\pm65$ & $8580\pm64$ & $8610\pm66$ & $8828\pm64$\\
$(1,1)$ & $^{3}D_{1}$ & $8054\pm66$ & $8286\pm64$ & $8347\pm65$ & $8572\pm64$ & $8602\pm66$ & $8820\pm64$\\
$(1,1)$ & $^{3}D_{2}$ & $8061\pm65$ & $8293\pm64$ & $8354\pm65$ & $8580\pm64$ & $8610\pm66$ & $8828\pm64$\\
$(1,1)$ & $^{3}D_{3}$ & $8066\pm65$ & $8298\pm64$ & $8360\pm65$ & $8585\pm63$ & $8615\pm65$ & $8833\pm64$\\
$(1,1)$ & $^{5}D_{0}$ & $8051\pm66$ & $8283\pm65$ & $8344\pm66$ & $8569\pm64$ & $8599\pm66$ & $8817\pm64$\\
$(1,1)$ & $^{5}D_{1}$ & $8050\pm66$ & $8282\pm65$ & $8344\pm66$ & $8569\pm64$ & $8599\pm66$ & $8817\pm64$\\
$(1,1)$ & $^{5}D_{2}$ & $8057\pm65$ & $8289\pm64$ & $8351\pm65$ & $8577\pm64$ & $8607\pm66$ & $8825\pm64$\\
$(1,1)$ & $^{5}D_{3}$ & $8062\pm65$ & $8294\pm64$ & $8356\pm65$ & $8582\pm64$ & $8612\pm65$ & $8830\pm64$\\
$(1,1)$ & $^{5}D_{4}$ & $8073\pm65$ & $8305\pm64$ & $8367\pm65$ & $8593\pm63$ & $8623\pm65$ & $8841\pm63$\\
\addlinespace[3pt]
\end{longtable}

% END INLINE tables/full_bqcq}

% BEGIN INLINE bibliography

% END INLINE bibliography
\end{document}